\documentclass[fleqn,number,preprint,,sort&compress,3p]{elsarticle}

\usepackage{amsmath, amssymb}
\usepackage{booktabs}
\usepackage{caption}
\usepackage[colorlinks=true]{hyperref}
    \usepackage[capitalize,nameinlink]{cleveref}
\usepackage{lipsum}
\usepackage{mathtools}
\usepackage{multirow}
\usepackage{nameref}
\usepackage{placeins}
\usepackage{subcaption}
\usepackage{xcolor}
\usepackage{xurl}

\newlength\subfiglength
\begin{document}

\begin{frontmatter}

    \title{Large radiation back-flux from Monte Carlo simulations
        of fusion neutron-material interactions\tnoteref{t1}}
    \tnotetext[t1]{This work was supported by the U.S. Department of Energy
        Office of Fusion Energy Sciences (DOE-FES) [grant number
        89233218CNA000001].}

    \author[1,2]{Michael A. Lively\corref{cor}}  
    \cortext[cor]{Corresponding author.}
    \ead{livelym@lanl.gov}

    \author[1]{Danny Perez}  
    \ead{danny_perez@lanl.gov}

    \author[3]{Blas P. Uberuaga}  
    \ead{blas@lanl.gov}

    \author[1]{Yanzeng Zhang}  
    \ead{yzengzhang@lanl.gov}

    \author[1]{Xian-Zhu Tang}  
    \ead{xtang@lanl.gov}

    \affiliation[1]{
        organization = {Theoretical Division, Los Alamos National Laboratory},
        city={Los Alamos},
        state={NM},
        postcode={87545},
        country={United States of America}
    }
    \affiliation[2]{
        organization = {X Computational Physics Division, Los Alamos National
            Laboratory},
        city={Los Alamos},
        state={NM},
        postcode={87545},
        country={United States of America}
    }
    \affiliation[3]{
        organization = {Materials Science and Technology Division, Los Alamos
            National Laboratory},
        city={Los Alamos},
        state={NM},
        postcode={87545},
        country={United States of America}
    }

    \begin{abstract}
        Fusion power reactors will generate intense neutron fluxes into
        plasma-facing and structural materials. Radiation back-fluxes,
        generated from neutron-material interactions under these fluxes, can
        dramatically impact the plasma dynamics, e.g., by seeding runaway
        electrons during disruptions via Compton scattering of background
        electrons by wall-emitted gamma radiation. Here, we quantify these
        back-fluxes, including neutrons, gamma rays, and electrons, using
        Monte Carlo calculations for a range of structural material candidates
        and first wall thicknesses. The radiation back-flux magnitudes are
        remarkably large, with neutron and gamma radiation back-fluxes on the
        same order of magnitude as the incident fusion neutron flux. Electron
        back-fluxes are two orders of magnitudes lower, but are emitted at
        sufficiently high energies to provide a relatively large back-current
        through the sheath which may cause sheath reversal. Material
        configuration plays a key role in determining back-flux magnitudes.
        The structural material chiefly determines the neutron back-flux
        magnitude, while the first wall thickness principally attenuates the
        gamma ray and electron back-fluxes. In addition to prompt back-fluxes,
        which are emitted immediately after fusion neutrons impact the
        surface, significant delayed gamma ray and electron back-fluxes arise
        from nuclear decay processes in the activated materials. These delayed
        back-flux magnitudes range from 2\%--7\% of the prompt back-fluxes,
        and remain  present during transients when fusion no longer occurs.
        During disruptions, build-up of delayed gamma radiation back-flux
        represents potential runaway electron seeding mechanisms, posing
        additional challenges for disruption mitigation in a power reactor
        compared with non-nuclear plasma operations. This work highlights the
        impact of these radiation back-fluxes plasma performance and
        demonstrates the importance of considering back-flux generation in
        materials selection for fusion power reactors.
    \end{abstract}


\end{frontmatter}

%
\section{Introduction}
\label{sec:intro}

In contrast to past and current research devices, burning-plasma operations in
future fusion power plants will generate large fluxes of 14.1-MeV
fusion neutrons. In ITER, which is expected to achieve 500 MW of
sustained fusion power over pulses of $\sim$400 s \cite{ITER-RP-2024}, the
rate of 14.1-MeV neutron production from D-T fusion would be
\begin{equation}
    \label{eq:fusion-rate}
    R_\mathrm{f} = \frac{P_\mathrm{f}}{E_\mathrm{f}}
    = 1.77\times 10^{20}~\text{s}^{-1}
\end{equation}
where $P_\mathrm{f}$ is the fusion power of 500 MW, and $E_\mathrm{f} =
17.6~\text{MeV}$ is the energy released per D-T fusion reaction. Given an
estimated plasma-facing surface area $A_\mathrm{w} \sim 660~\text{m}^2$, this
equates to an average neutron flux to the walls and divertor of
$\phi_\mathrm{n} \sim 2.68\times 10^{17}~\text{m}^{-2}~\text{s}^{-1}$. In
fusion power reactors with even higher power outputs, the fusion neutron
fluxes will approach magnitudes similar to those found in fission power
reactors ($\phi_\mathrm{n}\sim10^{18}\text{--}
10^{20}~\text{m}^{-2}~\text{s}^{-1}$), thus posing industrial-scale materials
challenges.

To date, fusion neutron-material interactions (NMI) have been investigated
from several perspectives. Active NMI research topics include material damage
and degradation under neutron irradiation \cite{Ishino-1996-JNuclMater,
    Mata-2011-FusEngDes, Rubel-2019-JFusEne,Spitsyn-2019-FusEngDes,
    Breidokaite-2023-RadPhysChem, Li-2024-JNuclMater}, tritium breeding
\cite{Hernandez-2018-FusEngDes, Rubel-2019-JFusEne, Segantin-2020-FusEngDes,
    Bae-2024-NuclFus, Kim-2024-FusEngDes, Prost-2024-JNuclMater},
    whole-facility
neutron transport for radiation safety \cite{Ghani-2015-FusEngDes,
    Wilson-2018-FusSciTech, Royston-2019-FusSciTech, Qiu-2024-FusEngDes}, and
radioactive waste generation \cite{El-Guebaly-2005-FusSciTech,
    Pampin-2012-FusEngDes, Someya-2017-FusSciTech, Gilbert-2018-FusEngDes,
    Cao-2021-FusEngDes, Bae-2024-NuclFus}. However, the potential impact of
    NMI on
the plasma performance, principally due to radioactive decay and emission from
activated first wall and structural materials, has not yet been investigated,
largely because such phenomena are not encountered in non-power-reactor
discharges. In this work, we aim to take the first steps towards understanding
and characterizing this impact as a basis for evaluating candidate fusion
reactor materials.

In practice, NMI in the first wall and structural materials will generate
secondary radiation which then transports back to the burning plasma, which we
term the \textit{radiation back-flux}. Radiation back-fluxes arise from two
major categories of mechanisms: prompt back-fluxes result directly from
nuclear and photoatomic collisions, while delayed back-fluxes are those
emitted from nuclear decay processes in activated materials. Both categories
of back-flux may impact the plasma through several mechanisms, including:
\begin{itemize}
    \item \textbf{Neutron} back-fluxes include secondary and reflected
    neutrons, which may cross the plasma volume to impact other plasma-facing
    surfaces and generate additional back-fluxes.
    \item \textbf{Gamma radiation} may seed runaway electrons by Compton
    scattering of plasma electrons to higher energies
    \cite{Martin-Solis-2017-NuclFus, Vallhagen-2020-JPP, Vallhagen-2024-NF}.
    \item \textbf{Electron emission} from photoatomic interactions
    and/or beta decay, with energies as high as a few MeV, can alter the
    dynamics of the plasma sheath. These can supply low energy electrons
    through secondary electron emission when the high energy electrons are
    brought back to rebombard the wall via electron gyroorbits. The presence
    of these electrons from the wall can modify the sheath potential,
    causing qualitative changes in the sheath dynamics
    \cite{Campanell-2013-PhysRevE, Campanell-2016-PhysRevLett,
        Bradshaw-2024-PSST}.
\end{itemize}

In this work, we carry out NMI simulations to characterize the emitted
radiation back-fluxes in a fusion power reactor. Furthermore, we elucidate the
effect of material design parameters including the first wall thickness and
choice of primary structural material. Our simulations show remarkably large
radiation back-fluxes, with neutron and gamma radiation back-fluxes in
particular being of the same order of magnitude as the incident fusion neutron
flux. The high-energy electron back-fluxes are two orders of magnitude lower,
but remain substantial enough to impact boundary plasma dynamics. These
results highlight the importance of coupled plasma physics/NMI simulations to
quantify the impact of the radiation back-fluxes on plasma performance during
both steady-state fusion power operations and transient (e.g., disruption)
conditions.

%
\section{Methods}
\label{sec:methods}

\begin{figure}[htb]
    \centering
    \includegraphics[width=\subfiglength]{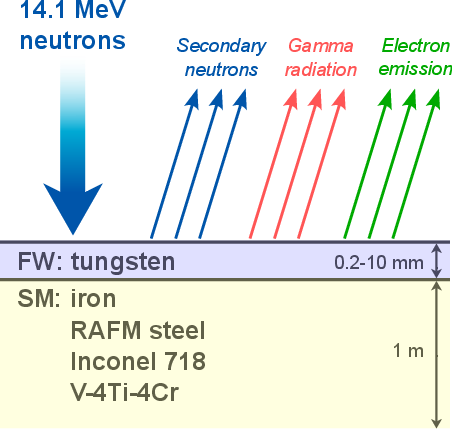}
    \caption{MCNP simulation setup for this work, consisting of 14.1-MeV
        neutrons incident on a first wall (FW) layer on top of a structural
        material (SM) layer. Note that arrow positions for the various
        back-fluxes
        is schematic only and does not represent an actual spatial
        distribution of
        back-fluxes.}
    \label{fig:mcnp-setup}
\end{figure}

For our NMI simulations, we use the Monte Carlo N-Particle (MCNP) code,
version 6.3.0 \cite{MCNP6-3}. MCNP is a general-purpose radiation transport
code which can handle most kinds of particles, including neutrons, photons,
electrons, and heavy charged particles. The general simulation setup is shown
in \cref{fig:mcnp-setup}. The target material system consists of a tungsten
first wall (FW) layer of variable thickness (0.2, 0.5, 2, or 10 mm) atop a
one-meter layer of a primary structural material (SM) candidate selected from
iron (for comparison purposes), reduced-activation ferritic-martensitic (RAFM)
steel \cite{Kohyama-1996-JNuclMater,
    Federici-2017-NuclFus}, Inconel 718 \cite{Bae-2022-NuclFus}, or vanadium
    alloy
\cite{Kurtz-2004-JNuclMater}. The material compositions
are given in Table \ref{tab:materials}. 14.1-MeV fusion neutrons enter the
FW surface at normal incidence. Nuclear data, which describes the complete
interactions between the various radiation types and materials, is taken from
the ENDF/B-VIII.0 library for neutrons \cite{Brown-2018-NDS,
    Conlin-2018-LAUR}, the EPRDATA14 library for photons
    \cite{Cullen-2014-EPICS,
    Hughes-2014-PNST, Hughes-2017-ICRS}, the EL03 library for electrons
\cite{Berger-1988-MCTrans, Seltzer-1988-MCTrans, Adams-2000-LAUR}, and the
ENDF7U library for photonuclear reactions (although these are quite rare in
practice) \cite{Chadwick-2006-NDS, Parsons-2022-LAUR}.

\begin{table}[h]
    \def\0{\hspace{\widthof{0}}}
    \centering
    \caption{Material compositions used for MCNP simulations. Elemental
        concentrations are given in percent atomic (at.\%).}
    \label{tab:materials}
    \begin{tabular}{l|ccccc}
        \toprule
        Element    & First wall & Iron & RAFM steel & Inconel 718 & V-4Ti-4Cr
        \\
        \midrule
        Carbon     &            &      & \00.1\0    & \00.08      &         \\
        Aluminum   &            &      &            & \00.50      &         \\
        Silicon    &            &      & \00.1\0    & \00.35      &         \\
        Titanium   &            &      &            & \00.90      & \04.0\0 \\
        Vanadium   &            &      & \00.25     &             & balance \\
        Chromium   &            &      & \08.5\0    & 19.0\0      & \04.0\0 \\
        Manganese  &            &      & \00.5\0    & \00.35      &         \\
        Iron       &            & 100  & balance    & balance     &         \\
        Cobalt     &            &      &            & \01.00      &         \\
        Nickel     &            &      &            & 52.0\0      &         \\
        Copper     &            &      &            & \00.30      &         \\
        Niobium    &            &      &            & \05.00      &         \\
        Molybdenum &            &      &            & \03.00      &         \\
        Tantalum   &            &      & \00.07     &             &         \\
        Tungsten   & 100        &      & \02.0\0    &             &         \\
        \bottomrule
    \end{tabular}
\end{table}

Output from the simulation consists of back-fluxes for neutrons, photons, and
electrons exiting the FW surface, which are tallied into bins according to the
exit energy and time (taking $t=0$ at incident neutron arrival). These
back-fluxes are normalized to be per incident neutron, and we adhere to this
convention in presenting the results throughout this paper. Each MCNP
simulation consist of $10^6$ incident neutron histories. Furthermore, we use
the \texttt{ACT} (activation) card to simulate delayed particle emission and
transport from nuclear decay of radioactive isotopes resulting from neutron
interactions. The full decay solver \cite{CINDER-90,Josey-2024-email} is quite
complex, but consists of two major steps: (1) time-integration of the system
of decay equations up to $10^{10} \text{s}$, and (2) sampling of delayed
particle emission from energy distributions given by model physics and/or
delayed emission library data. MCNP then simulates the transport of delayed
particles as requested by the user.

%
\section{Results}
\label{sec:results}

\cref{tab:summary} summarizes the various radiation back-fluxes for each
combination of FW thickness and SM selection presented in this work. The most
salient features of these results are that:
\begin{itemize}
    \item Fusion neutron irradiation of reactor materials generates
    significant radiation back-fluxes, particularly of neutrons and gamma
    rays. Neutron back-fluxes are of the same order of magnitude as the
    incident fusion neutron flux, while gamma radiation back-fluxes are within
    an order of magnitude of the incident flux.
    \item Material design parameters play a key role in determining the
    back-flux magnitudes. Neutron back-fluxes depend strongly on the choice of
    structural material, while the first wall thickness is the dominant factor
    influencing the gamma ray and electron back-fluxes.
    \item Delayed gamma ray and electron back-fluxes build up to significant
    levels during sustained fusion power operations. The terminal delayed
    gamma ray back-flux magnitude ranges from 2--7\% of the prompt back-flux
    magnitude. These back-flux levels remain substantial enough to influence
    plasma dynamics during transient, non-fusion events such as disruptions.
\end{itemize}
These key outcomes highlight the importance of our results and
demonstrate why radiation back-fluxes are a critical topic for fusion power
reactor design and operations. In the following sections, we emphasize
exemplary results from our simulations to demonstrate clearly each of these
points.

\begin{table}[ht]
    \def\0{\hspace{\widthof{0}}}
    \centering
    \caption{Summary of radiation back-fluxes from fusion neutron interactions
        with various material configurations. Fluxes are
        normalized to the incident neutron flux ($\phi_\mathrm{n}$).
        $\phi_\mathrm{n'}$: total neutron back-flux.
        $\phi_{\gamma,\mathrm{p}},~\phi_\mathrm{e,p}$: prompt gamma ray and
        electron back-fluxes. $\phi_{\gamma,\mathrm{d}},~\phi_\mathrm{e,d}$:
        terminal delayed gamma ray and electron back-fluxes.
        $\phi_{\gamma,\mathrm{tot}},~\phi_\mathrm{e,tot}$: total gamma ray and
        electron back-fluxes.}
    \label{tab:summary}
    \begin{tabular}{l|cccc}
        \toprule
        Iron & FW  = 0.2 mm & FW = 0.5 mm & FW = 2 mm & FW = 10 mm \\
        \midrule
        $\displaystyle \phi_\mathrm{n'}$
        & 0.8908 & 0.8925 & 0.9006 & 0.9419 \\[1em]
        $\displaystyle \phi_{\gamma,\mathrm{p}}$
        & 0.5075 & 0.4667 & 0.3782 & 0.2497 \\
        $\displaystyle \phi_{\gamma,\mathrm{d}}$
        & 0.0165 & 0.0147 & 0.0102 & 0.0032\\
        $\displaystyle \phi_{\gamma,\mathrm{tot}}$
        & 0.5239 & 0.4813 & 0.3884 & 0.2530 \\[1em]
        $\displaystyle \phi_\mathrm{e,p}$
        & $5.65\times 10^{-3}$ & $5.58\times 10^{-3}$
        & $4.49\times 10^{-3}$ & $2.58\times 10^{-3}$ \\
        $\displaystyle \phi_\mathrm{e,d}$
        & $1.30\times 10^{-4}$ & $1.14\times 10^{-4}$
        & $8.6\0\times 10^{-5}$ & $4.7\0\times 10^{-5}$ \\
        $\displaystyle \phi_\mathrm{e,tot}$
        & $5.78\times 10^{-3}$ & $5.69\times 10^{-3}$
        & $4.58\times 10^{-3}$ & $2.63\times 10^{-3}$ \\
        \bottomrule
        \multicolumn{5}{c}{\-} \\
        \toprule
        RAFM steel & FW  = 0.2 mm & FW = 0.5 mm & FW = 2 mm & FW = 10 mm \\
        \midrule
        $\displaystyle \phi_\mathrm{n'}$
        & 0.8458 & 0.8480 & 0.8578 & 0.9050 \\[1em]
        $\displaystyle \phi_{\gamma,\mathrm{p}}$
        & 0.4895 & 0.4549 & 0.3742 & 0.2491 \\
        $\displaystyle \phi_{\gamma,\mathrm{d}}$
        & 0.0150 & 0.0134 & 0.0094 & 0.0030 \\
        $\displaystyle \phi_{\gamma,\mathrm{tot}}$
        & 0.5045 & 0.4683 & 0.3836 & 0.2521 \\[1em]
        $\displaystyle \phi_\mathrm{e,p}$
        & $5.81\times 10^{-3}$ & $5.51\times 10^{-3}$
        & $4.31\times 10^{-3}$ & $2.68\times 10^{-3}$ \\
        $\displaystyle \phi_\mathrm{e,d}$
        & $1.17\times 10^{-4}$ & $1.06\times 10^{-4}$
        & $7.4\0\times 10^{-5}$ & $3.1\0\times 10^{-5}$ \\
        $\displaystyle \phi_\mathrm{e,tot}$
        & $5.92\times 10^{-3}$ & $5.62\times 10^{-3}$
        & $4.38\times 10^{-3}$ & $2.71\times 10^{-3}$ \\
        \bottomrule
        \multicolumn{5}{c}{\-} \\
        \toprule
        Inconel 718 & FW  = 0.2 mm & FW = 0.5 mm & FW = 2 mm & FW = 10 mm \\
        \midrule
        $\displaystyle \phi_\mathrm{n'}$
        & 0.6292 & 0.6327 & 0.6492 & 0.7301 \\[1em]
        $\displaystyle \phi_{\gamma,\mathrm{p}}$
        & 0.4245 & 0.3958 & 0.3298 & 0.2337 \\
        $\displaystyle \phi_{\gamma,\mathrm{d}}$
        & 0.0290 & 0.0250 & 0.0165& 0.0044 \\
        $\displaystyle \phi_{\gamma,\mathrm{tot}}$
        & 0.4536 & 0.4209 & 0.3464 & 0.2381 \\[1em]
        $\displaystyle \phi_\mathrm{e,p}$
        & $5.19\times 10^{-3}$ & $4.83\times 10^{-3}$
        & $3.87\times 10^{-3}$ & $2.50\times 10^{-3}$ \\
        $\displaystyle \phi_\mathrm{e,d}$
        & $1.93\times 10^{-4}$ & $1.66\times 10^{-4}$
        & $1.05\times 10^{-4}$& $4.6\0\times 10^{-5}$ \\
        $\displaystyle \phi_\mathrm{e,tot}$
        & $5.38\times 10^{-3}$ & $4.99\times 10^{-3}$
        & $3.98\times 10^{-3}$ & $2.55\times 10^{-3}$ \\
        \bottomrule
        \multicolumn{5}{c}{\-} \\
        \toprule
        V-4Ti-4Cr & FW  = 0.2 mm & FW = 0.5 mm & FW = 2 mm & FW = 10 mm \\
        \midrule
        $\displaystyle \phi_\mathrm{n'}$
        & 0.9132 & 0.9137 & 0.9184 & 0.9456 \\[1em]
        $\displaystyle \phi_{\gamma,\mathrm{p}}$
        & 0.4504 & 0.4063 & 0.3323 & 0.2532 \\
        $\displaystyle \phi_{\gamma,\mathrm{d}}$
        & 0.0133 & 0.0117 & 0.0084 & 0.0036 \\
        $\displaystyle \phi_{\gamma,\mathrm{tot}}$
        & 0.4637 & 0.4181 & 0.3407 & 0.2568 \\[1em]
        $\displaystyle \phi_\mathrm{e,p}$
        & $6.08\times 10^{-3}$ & $5.88\times 10^{-3}$
        & $4.85\times 10^{-3}$ & $2.94\times 10^{-3}$ \\
        $\displaystyle \phi_\mathrm{e,d}$
        & $1.15\times 10^{-4}$ & $9.8\0\times 10^{-5}$
        & $7.0\0\times 10^{-5}$ & $4.0\0\times 10^{-5}$ \\
        $\displaystyle \phi_\mathrm{e,tot}$
        & $6.20\times 10^{-3}$ & $5.98\times 10^{-3}$
        & $4.92\times 10^{-3}$ & $2.98\times 10^{-3}$ \\
        \bottomrule
    \end{tabular}
\end{table}

\FloatBarrier

\subsection{Prompt radiation back-fluxes in fusion power reactors}
\label{subsec:prompt-flux}

\begin{figure}[htb]
    \centering
    \begin{subfigure}{0.475\linewidth}
        \includegraphics[width=\linewidth]{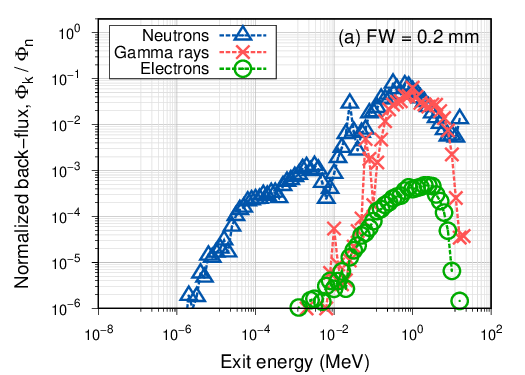}
        \centering
        \phantomcaption
        \label{subfig:erg-rafm0.2}
    \end{subfigure}
    \begin{subfigure}{0.475\linewidth}
        \includegraphics[width=\linewidth]{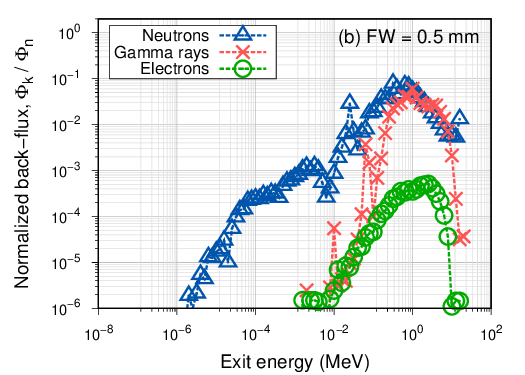}
        \centering
        \phantomcaption
        \label{subfig:erg-rafm0.5}
    \end{subfigure}
    \begin{subfigure}{0.475\linewidth}
        \includegraphics[width=\linewidth]{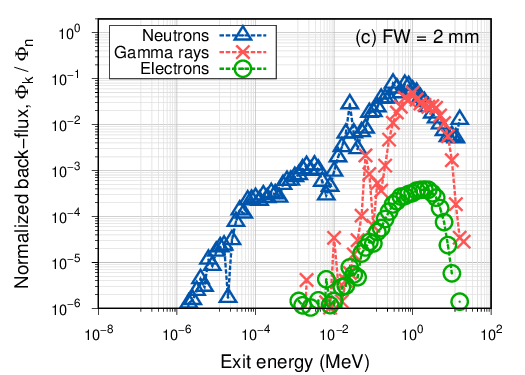}
        \centering
        \phantomcaption
        \label{subfig:erg-rafm2}
    \end{subfigure}
    \begin{subfigure}{0.475\linewidth}
        \includegraphics[width=\linewidth]{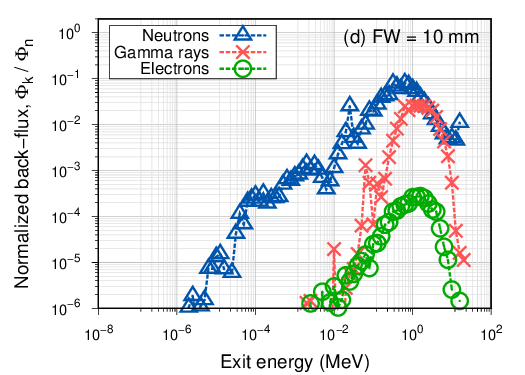}
        \centering
        \phantomcaption
        \label{subfig:erg-rafm10}
    \end{subfigure}
    \caption{Energy-resolved, time-integrated radiation back-fluxes of
        neutrons, gamma rays, and relativistic electrons for RAFM steel
        structural material with various combinations of tungsten first wall
        thickness (0.2, 0.5, 2, or 10 mm).}
    \label{fig:backflux-rafm}
\end{figure}

\cref{fig:backflux-rafm} shows radiation back-flux energy distributions for
different first wall (FW) thicknesses with RAFM steel as the exemplary
structural material (SM). Similar distributions for other considered structural
materials are given in the Supplementary Material. Nearly all of
the back-fluxes in each plot occur within $\Delta t \le 0.1~\text{ms}$ and is
therefore considered prompt. All neutron back-fluxes are prompt (i.e., there
is no delayed neutron emission\footnote{This is very different from the case
    in a nuclear \textit{fission} reactor, in which delayed neutron emission is
    not only present but plays a key role in the reactor kinetics.}), while for
gamma radiation and electron emission the delayed back-fluxes contribute only
a few percent of the total magnitude as seen from \cref{tab:summary}. As a
rule, this is true for all SM candidates studied in this work. The discussion
immediately following therefore chiefly addresses the prompt back-fluxes,
while delayed back-fluxes are discussed in \cref{subsec:delayed-flux}.

\begin{figure}[htb]
    \centering
    \begin{subfigure}{0.475\linewidth}
        \includegraphics[width=\linewidth]{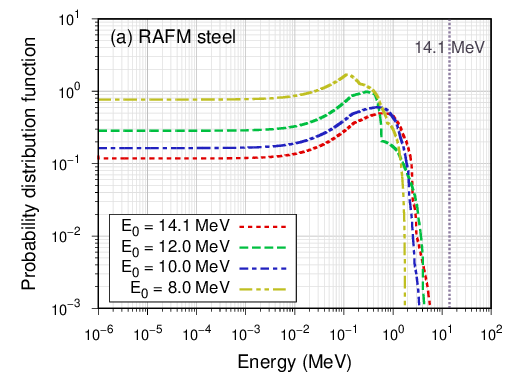}
        \centering
        \phantomcaption
        \label{subfig:sec-erg-rafm}
    \end{subfigure}
    \begin{subfigure}{0.475\linewidth}
        \includegraphics[width=\linewidth]{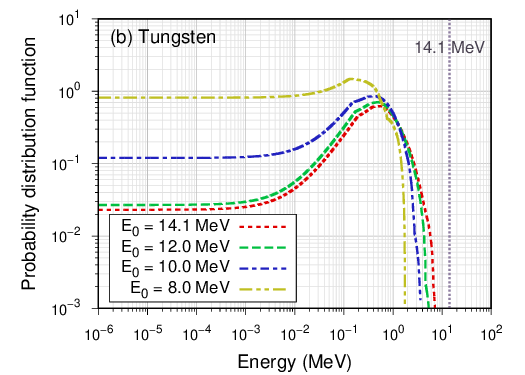}
        \centering
        \phantomcaption
        \label{subfig:sec-erg-w}
    \end{subfigure}
    \caption{Energy distributions of secondary neutrons from spallation
        collisions in (\subref{subfig:sec-erg-rafm}) RAFM steel and
        (\subref{subfig:sec-erg-w}) tungsten, extracted from ENDF/B-VIII.0
        cross
        section tabular data files \cite{Conlin-2018-LAUR}. Dashed vertical
        lines
        indicate the 14.1 MeV fusion neutron energy as a reference point.}
    \label{fig:n-sec-erg}
\end{figure}

Out of the three radiation types considered, the neutron back-flux is
consistently the largest in all cases, ranging from 63\% to 95\% of the
incident fusion neutron flux magnitude. Physically, the neutron back-flux
arises from several mechanisms. The high-energy peak between 10 and 14 MeV
results from backscattering of incident neutrons by one or a few elastic
collisions. The largest, broad peak spanning from $\sim$50 keV up to $\sim$10
MeV correlates with the energy distribution of secondary neutrons from
spallation collisions (that is, collisions leading to ejection of additional
neutrons and/or other nuclear fragments). \cref{fig:n-sec-erg} shows examples
of these secondary neutron energy distributions extracted from ENDF/B-VIII.0
cross section tables for RAFM steel and tungsten. The peaks of these
distributions match well with the broad peaks in the neutron back-flux
distributions. At lower energies, around 25 keV, the back-flux distributions
show a sharp peak which is specific to the selected SM and most likely
corresponds to a nuclear resonance of Fe-56. The same peak appears when pure
iron is the SM, while different, smaller peaks appear when Inconel 718 (around
10 keV) or V-4Ti-4Cr (around 60 keV) are the SM. Finally, the neutron
back-flux distribution tails off below energies of several keV, a region
corresponding to low-energy secondary and reflected neutrons which diffuse out
of the wall.

\begin{figure}[htb]
    \centering
    \includegraphics[width=0.475\linewidth]{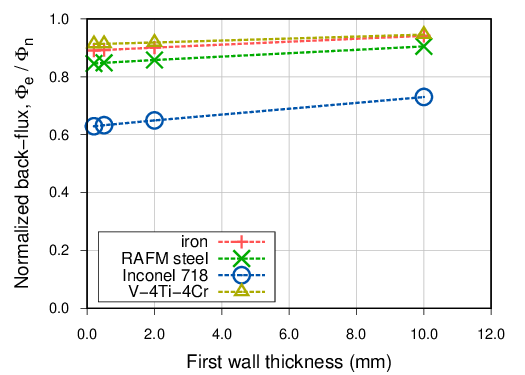}
    \caption{Total neutron back-fluxes summed over all angular and time bins
        for various combinations of FW thickness and SM selection.}
    \label{fig:backflux-nn}
\end{figure}

\cref{fig:backflux-nn} shows the total neutron back-fluxes for each simulated
combination of FW thickness and SM selection. The effect of FW thickness is
minor, as for each SM choice the back-flux magnitude increases by $\sim$10\%
as FW thickness increases from 0.2 mm to 10 mm. This is due to the greater
neutron capture and scattering cross-sections of tungsten compared to the
smaller nuclei making up the structural materials. The effect of SM choice is
much stronger, as the nickel-based Inconel 718 alloy SM leads to 25--30\%
lower neutron back-fluxes than the iron, RAFM steel, or vanadium alloy SMs.
The strong effect of SM and the weak effect of FW thickness indicate that most
neutron scattering and multiplication takes place in the SM rather than the FW.

\begin{figure}[htb]
    \centering
    \begin{subfigure}{0.475\linewidth}
        \centering
        \includegraphics[width=\linewidth]{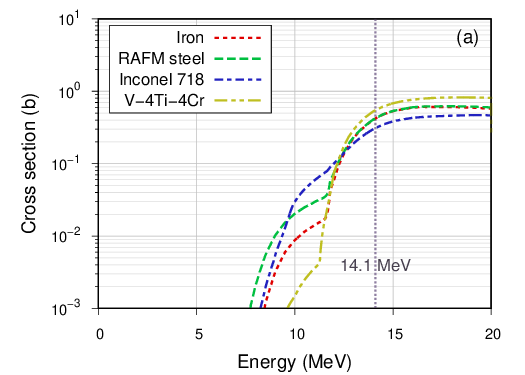}
        \phantomcaption
        \label{subfig:sm-nmult}
    \end{subfigure}
    \begin{subfigure}{0.475\linewidth}
        \centering
        \includegraphics[width=\linewidth]{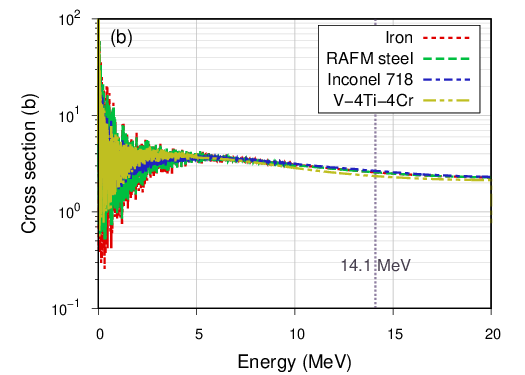}
        \phantomcaption
        \label{subfig:sm-xs-tot}
    \end{subfigure}
    \caption{Important cross sections for neutron transport in structural
        materials, extracted from ENDF/B-VIII.0 cross section tabular data
        files
        \cite{Conlin-2018-LAUR}: (\subref{subfig:sm-nmult}) neutron
        multiplication,
        (\subref{subfig:sm-xs-tot}) total cross section. Dashed vertical lines
        indicate the 14.1 MeV fusion neutron energy as a reference point.}
    \label{fig:sm-xs}
\end{figure}

\cref{subfig:sm-nmult} shows neutron multiplication cross sections for each SM
option, again extracted from the ENDF/B-VIII.0 cross section tabular data
files. Based on this, Inconel 718 most likely exhibits lower neutron
back-fluxes due to a smaller neutron multiplication cross section at the
incident energy of 14.1 MeV. Given this, since V-4Ti-4Cr has an even higher
multiplication cross section at the same energy, it is reasonable to ask why
the neutron back-flux with the vanadium alloy SM selected is only slightly
larger than that for iron-based SMs. To address this question,
\cref{subfig:sm-xs-tot} shows the total cross sections for each SM option in
the same energy range. V-4Ti-4Cr has a lower total cross section than the
other SM options between 10 and 14.1 MeV, which means that high-energy
neutrons will penetrate more deeply into the vanadium-based material before a
collision. On the other hand, for neutron energies of about 2--5 MeV the total
cross section in V-4Ti-4Cr is larger on average than that of the other SM
options, which means that secondary neutrons emitted in this energy range are
more likely to be stopped inside the material. These factors balance out
the larger neutron multiplication rate of the vanadium alloy, resulting in a
neutron back-flux which is only modestly larger than that for iron-based SMs.

\begin{figure}[htb]
    \centering
    \begin{subfigure}{0.475\linewidth}
        \centering
        \includegraphics[width=\linewidth]{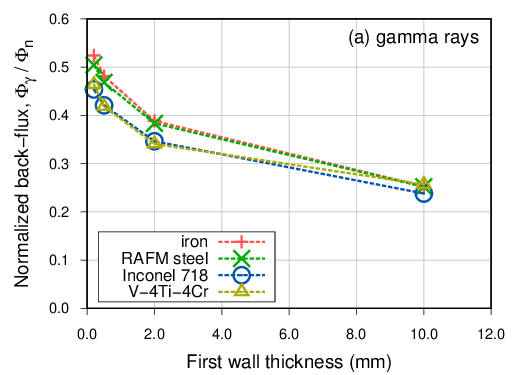}
        \phantomcaption
        \label{subfig:backflux-ph}
    \end{subfigure}
    \begin{subfigure}{0.475\linewidth}
        \centering
        \includegraphics[width=\linewidth]{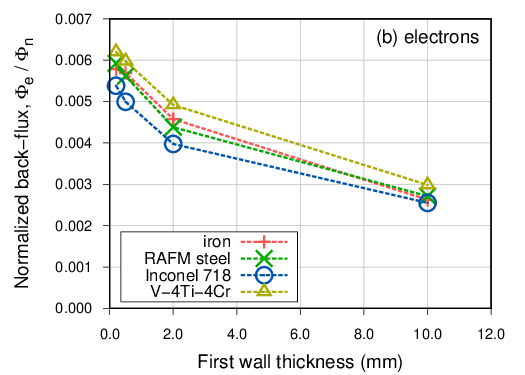}
        \phantomcaption
        \label{subfig:backflux-el}
    \end{subfigure}
    \caption{Total (\subref{subfig:backflux-ph}) gamma and
        (\subref{subfig:backflux-el}) electron radiation back-fluxes out of the
        simulated wall summed over all time and energy bins.}
    \label{fig:backflux-ph-el}
\end{figure}

Considering the other radiation types, summarized in
\cref{fig:backflux-ph-el}, gamma radiation back-fluxes are
typically less than the neutron back-fluxes, but within an order of magnitude
of the incident fusion neutron flux (see \cref{tab:summary}). Electron emission
back-fluxes are consistently around two orders of magnitude lower than the
gamma radiation back-fluxes. Aside from the disparity in magnitudes, gamma
radiation and electron emission back-fluxes follow the same trends. This is
what we expect: since the primary method for (prompt) electron production is
from photoatomic interactions, gamma radiation and electron emission should be
closely correlated.

The trends with material parameters for gamma radiation and electron emission
back-fluxes are somewhat opposite to the trends for neutron back-fluxes.
Specifically, the effect of the SM selection is quite small. On the other
hand, the effect of the FW thickness is much more prominent, as the back-flux
magnitudes decrease by $\sim$50\% as the FW thickness increases from 0.2 mm to
10 mm. This indicates that the gamma radiation back-flux originates primarily
in the SM and is attenuated by passing through the FW, since tungsten with a
high atomic number of $Z=74$ has much higher photon and electron stopping
powers than any alloys based on the first row of transition metals ($Z=22$ to
$Z=29$).

\subsection{Delayed radiation back-fluxes in steady-state fusion power
    reactors}
\label{subsec:delayed-flux}

The delayed back-fluxes, broadly defined to consist of all back-fluxes exiting
the wall for $t > 0.1~\text{ms}$, make up only a few percent of the total
gamma ray and electron back-fluxes as seen in \cref{tab:summary} and may be
widely distributed over times up to the reactor lifetime ($\sim$30 years).
Therefore, at first glance the delayed contributions may seem negligible---this
is not the case in reality. In fact, because the delayed back-fluxes are
comparatively long-lived, they can build up to potentially significant levels
over the course of a fusion reactor's lifetime. In this section, we show the
buildup of delayed back-fluxes and the magnitudes which may be reached over
the reactor lifetime.

(Note: we must emphasize that \textit{there is no delayed back-flux of
    neutrons} in any of our simulations, because none of the transmutation
products decay by neutron emission. Only the gamma radiation and electron
emission back-fluxes include delayed contributions.)

\begin{figure}[htb]
    \centering
    \begin{subfigure}{0.475\linewidth}
        \centering
        \includegraphics[width=\linewidth]{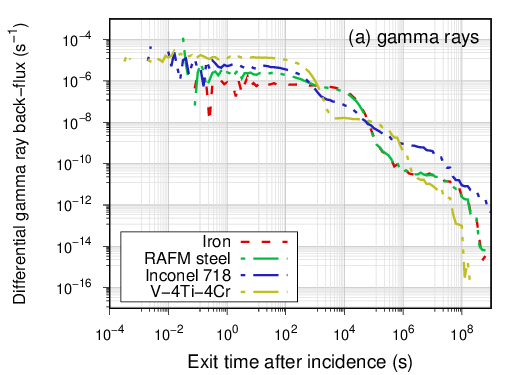}
        \phantomcaption
        \label{subfig:delay-flux-ph}
    \end{subfigure}
    \begin{subfigure}{0.475\linewidth}
        \centering
        \includegraphics[width=\linewidth]{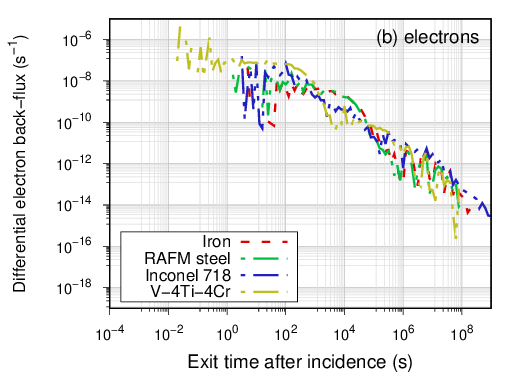}
        \phantomcaption
        \label{subfig:delay-flux-el}
    \end{subfigure}
    \caption{Time-differential delayed back-fluxes, $\nu(\Delta t - t')$, of
        (\subref{subfig:delay-flux-ph}) gamma rays and
        (\subref{subfig:delay-flux-el}) electrons for different structural
        materials with a 2-mm first wall thickness.}
    \label{fig:delay-flux}
\end{figure}

To begin, let $\nu(\Delta t - t')$ represent the differential delayed
back-flux, with dimensions of $t^{-1}$ per incident neutron, emitted at time
$\Delta t$ due to a neutron incident at time $0 \le t' < \Delta t$ where
fusion power operations began at $t = 0$. \cref{fig:delay-flux} shows examples
of these distributions for delayed gamma rays and electrons for a FW thickness
of 2 mm in each case. By assuming steady-state fusion power operations with a
constant fusion neutron flux of $\phi_\mathrm{n}$, we may compute the total
delayed back-flux buildup from all incident neutrons during $0 \le t \le
\Delta t - t_\mathrm{c}$, where $t_\mathrm{c} = 0.1~\text{ms}$ is the cutoff
time separating prompt and delayed back-fluxes, as
\begin{equation}
    \label{eq:delayed-backflux}
    \phi_\mathrm{d} = \phi_\mathrm{n} \int_{0}^{\Delta t - t_\mathrm{c}}
    \nu(\Delta t - t')~dt'
\end{equation}
We may further consider the worst-case scenario, which is the terminal case as
$\Delta t\rightarrow\infty$ describing the back-fluxes for times beyond the
half-life of the longest-lived significant radioisotope. We note that while
\cref{eq:delayed-backflux} assumes a constant $\phi_\mathrm{n}$ for $0 \le t
\le \Delta t - t_\mathrm{c}$, a real steady-state fusion power reactor will
have down time for maintenance, disruption recovery, etc. This can be
represented by multiplying \cref{eq:delayed-backflux} by a duty factor
parameter, $0 < f_\mathrm{dc} < 1$ if required. As-is, then,
\cref{eq:delayed-backflux} represents the worst-case scenario for delayed
back-flux magnitudes with undisrupted steady-state operations during $0 \le t
\le \Delta t - t_\mathrm{c}$.

\begin{figure}[htb]
    \centering
    \begin{subfigure}{0.475\linewidth}
        \centering
        \includegraphics[width=\linewidth]{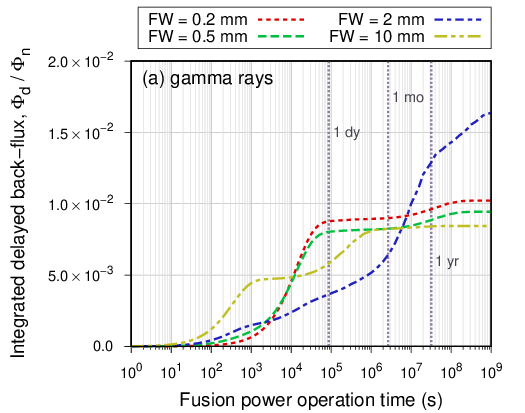}
        \phantomcaption
        \label{subfig:delay-int-ph}
    \end{subfigure}
    \begin{subfigure}{0.475\linewidth}
        \centering
        \includegraphics[width=\linewidth]{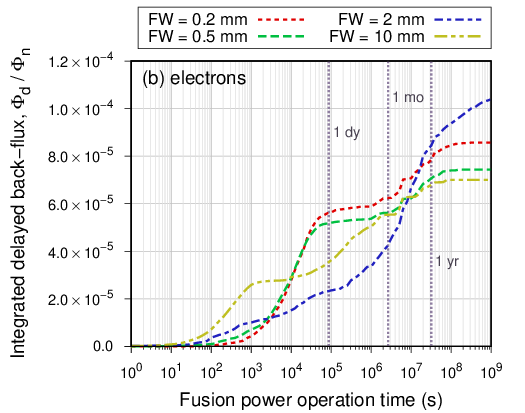}
        \phantomcaption
        \label{subfig:delay-int-el}
    \end{subfigure}
    \caption{Time-integrated delayed back-fluxes of
        (\subref{subfig:delay-int-ph}) gamma rays and
        (\subref{subfig:delay-int-el}) electrons for different
        structural materials with a 2-mm first wall
        thickness.}
    \label{fig:delay-int}
\end{figure}

\cref{fig:delay-int} shows the integrated back-fluxes of gamma rays and
electrons, per incident neutron, for each of the four simulated SMs with a FW
thickness of 2 mm in each case. It is immediately apparent that the electron
back-fluxes follow the same trends as the gamma ray back-fluxes except for
being more than two orders of magnitude smaller. Therefore, in the following
discussion the gamma ray and electron delayed back-fluxes are treated together.

The results in \cref{fig:delay-int} can be assessed for each SM selection:
\begin{itemize}
    \item With iron as the SM selection, a reactor will reach 86.7\% of the
    terminal back-flux level after 1 day of steady-state fusion power
    operations. The back-fluxes reach a terminal steady state within 20 years,
    which is within the nominal reactor lifetime of $\sim$30 years. The
    terminal delayed gamma radiation back-flux magnitude is
    $0.0102\phi_\mathrm{n}$, which is 2.7\% of the prompt back-flux magnitude.
    \item With RAFM steel as the SM, the behavior is almost identical to that
    with iron as the SM, except that using RAFM steel leads to a terminal
    gamma radiation back-flux magnitude of only $9.4\times
    10^{-3}\phi_\mathrm{n}$, a 7.8\% reduction compared to pure iron. This is
    an expected result, given the reduced-activation nature of RAFM steel, and
    is attributable to the presence of isotopes which, when activated, form
    radioisotopes with shorter half-lives than those formed from transmutation
    of pure iron. The terminal delayed gamma radiation back-flux magnitude in
    this case is 2.5\% of the prompt back-flux magnitude.
    \item With Inconel 718 as the SM, the comparison with iron-based materials
    initially looks favorable, as the built-up delayed back-flux magnitude is
    less than seen from using RAFM steel as the SM for the first 3--4 months
    of steady-state operations. Beyond this, however, the delayed back-flux
    magnitude with Inconel 718 is much greater, in large part because no
    terminal steady state is reached during the nominal 30-year reactor
    lifetime. The terminal delayed gamma radiation back-flux magnitude in
    this case is $0.0165\phi_\mathrm{n}$, which is 5.0\% of the prompt
    back-flux magnitude. This fraction is much larger than for any other SM
    selection.
    \item With V-4Ti-4Cr as the SM, the built-up delayed back-flux is
    significantly larger for the first few hours of reactor operation, but
    becomes lower than for the iron-based SM choices after about 3 hours. The
    built-up delayed back-flux reaches 98.6\% of the terminal magnitude after
    1 month of operations. The terminal delayed gamma radiation back-flux is
    $8.4\times 10^{-3}\phi_\mathrm{n}$, the lowest level for any SM option in
    this work and 2.5\% of the prompt back-flux magnitude.
\end{itemize}

\begin{figure}[htb]
    \centering
    \begin{subfigure}{0.475\linewidth}
        \centering
        \includegraphics[width=\linewidth]{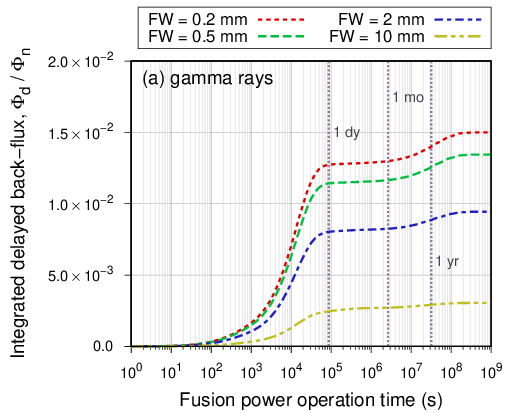}
        \phantomcaption
        \label{subfig:rafm-int-ph}
    \end{subfigure}
    \begin{subfigure}{0.475\linewidth}
        \centering
        \includegraphics[width=\linewidth]{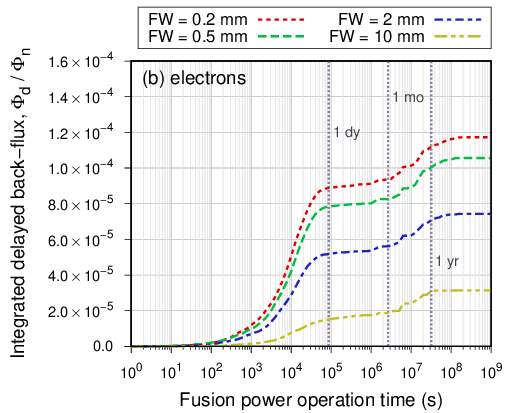}
        \phantomcaption
        \label{subfig:rafm-int-el}
    \end{subfigure}
    \caption{Time-integrated delayed back-fluxes of
        (\subref{subfig:rafm-int-ph}) gamma rays and
        (\subref{subfig:rafm-int-el}) electrons for different first wall
        thicknesses with RAFM steel structural material.}
    \label{fig:rafm-int}
\end{figure}

\cref{fig:rafm-int} shows the integrated back-fluxes of gamma rays and
electrons, per incident neutron, for each of the four simulated FW thicknesses
with RAFM steel as the exemplary SM in each case. As was the case for the
prompt back-fluxes (c.f. \cref{fig:backflux-ph-el}), the FW thickness has a
strong attenuation effect. In fact, the FW attenuation is modestly stronger
for the delayed back-fluxes than for the prompt ones, as the gamma ray and
electron terminal delayed back-fluxes decrease by $\sim$80\% as the FW
thickness increases from 0.2 mm to 10 mm, compared to a $\sim$50\% reduction
of the prompt back-fluxes over the same domain. This aside, the overall shape
of the integrated delayed back-flux curves with increasing fusion operations
time is not significantly affected by the FW thickness. This indicates that the
delayed back-fluxes originate primarily from decay processes in the SM rather
than in the FW, which agrees with the similar conclusion about neutron
multiplication drawn from the data in \cref{fig:backflux-nn}.

While even the largest built-up delayed back-flux level found from these
simulations is still $<7\%$ of the prompt back-flux level, this few-percent
magnitude is significant enough to influence the plasma dynamics under
transient conditions when fusion in the plasma stops. In these situations, the
prompt back-fluxes disappear but the delayed back-fluxes will remain for a
long time. These cases include transients such as reactor startup, shutdown,
and perhaps most importantly during major disruptions. Most concerning, the
build-up of delayed gamma radiation back-flux can drive runaway electron
seeding during disruptions via Compton scattering of background electrons,
which we discuss below in \cref{subsec:compton-seed}.

%
\section{Discussion}
\label{sec:discussion}

\subsection{Neutron multiplication impacts}
\label{subsec:n-mult}

The large neutron back-flux, which is of the same order of magnitude as the
incident fusion neutron flux, will transport across the plasma volume and, in
most cases, impact on other plasma-facing surfaces. This in turn leads to
additional gamma radiation and electron emission back-fluxes caused by these
secondary and reflected neutrons (though note that we expect very little to no
additional neutron back-flux from this phenomenon, since only 2--3\% of the
back-flux neutrons are above the threshold energy for neutron multiplication
reactions). To characterize the increased back-flux magnitudes, we conducted
an additional simulation with a SM of RAFM steel and a FW thickness of 2 mm,
in which we added a reflecting boundary for neutrons exiting the FW surface to
mimic transport of the neutron back-flux arriving from other hypothetical
plasma-facing surfaces. From this simulation, the total gamma radiation
back-flux magnitude is $\phi_{\gamma,\mathrm{tot}} = 0.5431\phi_\mathrm{n}$
(an additional 42\% over the simulation without a reflecting boundary) and the
electron emission back-flux magnitude is $\phi_\mathrm{e,tot} = 6.78\times
10^{-3}\phi_\mathrm{n}$ (an additional 55\%). As both of these magnitudes are
less than the neutron back-flux (85.8\% of incident), a conservative estimate
of the additional gamma radiation and electron back-flux magnitudes could be
made by multiplying values of $\phi_{\gamma,\mathrm{tot}}$ and
$\phi_\mathrm{e,tot}$ in \cref{tab:summary} by $(\phi_\mathrm{n} +
\phi_\mathrm{n'}) / \phi_\mathrm{n}$ in lieu of performing every simulation
with the more expensive reflection boundary condition. Of course, in a real
fusion reactor the actual back-flux magnitudes will depend on the device
geometry.

We also note that many generated secondary neutrons are directed into the
material rather than outward as back-flux. Therefore, in a real reactor
geometry the flux of secondary neutrons can contribute to the overall neutron
flux which reaches the tritium breeding blanket (TBB). Currently, many
material candidates have been proposed as neutron multipliers for the TBB
\cite{Hernandez-2018-FusEngDes}, including beryllium-based materials such as
TiBe\textsubscript{12} \cite{Kawamura-2003-NuclFus, Boccaccini-2022-FusEngDes,
    Mukai-2023-NuclMatEne, Zhou-2023-Energies}, VBe\textsubscript{12}
\cite{Mukai-2023-NuclMatEne, Nakamichi-2018-NuclMatEne}, or
CrBe\textsubscript{12} \cite{Zhou-2023-Energies}, and Pb-based materials such
as Zr\textsubscript{5}Pb\textsubscript{3} \cite{Gohar-1980-ANS,
    Donne-1986-JNuclMater}, LaPb\textsubscript{3}
    \cite{Gaisin-2023-JMatResTech},
Li\textsubscript{2}Pb\textsubscript{x}Ti\textsubscript{1-x}O\textsubscript{3}
\cite{Gao-2024-JNuclMater}. Choosing the multiplier material which has the
optimal combination of thermal, mechanical, and neutronics properties is a
daunting challenge for fusion power reactors. The secondary neutron flux
towards the TBB should be included in the analysis of these materials, which
may allow neutronics requirements to be satisfied more easily and a material
with excellent thermal and mechanical properties to be chosen. On the other
hand, significant neutron multiplication in non-breeder materials also holds
the potential for increased damage production in those materials. Both of
these factors should be key considerations in future studied motivated by the
present work.

\subsection{Compton scattering runaway electron seed from delayed gamma ray
    back-fluxes}
\label{subsec:compton-seed}

The delayed back-fluxes persist in a fusion power reactor even when fusion is
no longer occurring. This is most concerning during a major disruption, when
the delayed gamma radiation back-flux can induce a perpetual runaway electron
seed through Compton scattering of cold background electrons to high energies.
Compton scattering is proposed to be the dominant runaway reseeding mechanism
under conditions in which the critical energy, $W_\mathrm{c}$, for electrons to
runaway is high ($W_\mathrm{c} \gtrsim 10~\text{keV}$). This can be the case,
e.g., after high-Z impurity injection \cite{Rosenbluth-1997-NuclFus,
    Hesslow-2019-NuclFus, Vallhagen-2020-JPP}, a commonly proposed disruption
mitigation method for the thermal quench. While estimates of this Compton seed
current have been made
previously \cite{Martin-Solis-2017-NuclFus, Vallhagen-2020-JPP,
    Vallhagen-2024-NF}, here we carry out a similar estimate to demonstrate the
utility of our results to characterize the Compton seed current in terms of
the material configuration of a fusion reactor, thus aiding disruption
mitigation through materials design and engineering.

While plasma physics simulations of the runaway dynamics are necessary to
accurately predict the runaway current induced by Compton scattering, here we
can estimate the runaway seed current,
\begin{equation}
    \label{eq:seed-current}
    I_\mathrm{seed} = \dot{n}_\mathrm{re} A_\mathrm{xs} \tau_\mathrm{res} e
    \langle v_\mathrm{re}\rangle f(\xi_\mathrm{c})
\end{equation}
where $\dot{n}_\mathrm{re}$ is the runaway generation rate by Compton
scattering, $A_\mathrm{xs}$ is the cross-sectional area of the plasma volume,
$\tau_\mathrm{res}$ is the exponential current quench time constant following
\cite{Martin-Solis-2017-NuclFus}, $e$ is the elementary charge, $\langle
v_\mathrm{re}\rangle$ is the average speed of a high-energy electron produced
from Compton scattering, and $f(\xi_\mathrm{c})$ is the fraction of Compton
scattering electrons with directional cosines $\xi > \xi_\mathrm{c}$ relative
to the magnetic field direction (electrons outside of this range will radiate
their energy away and will not become runaways). The critical cosine value is
defined as $\xi_\mathrm{c} = \sqrt{2a/R_0}$ with $a$ and $R_0$ the minor and
major tokamak radii, respectively.

\begin{figure}[htb]
    \centering
    \includegraphics[width=0.475\linewidth]{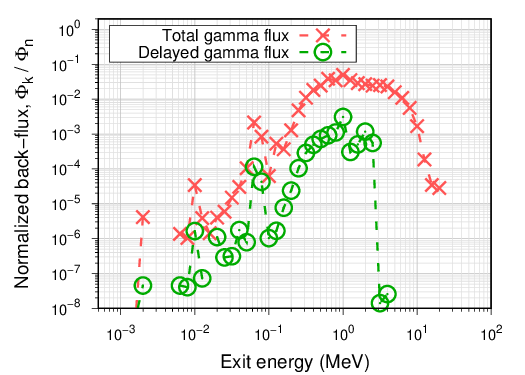}
    \caption{Energy distributions of the total and delayed gamma radiation
        back-fluxes for RAFM steel with a first wall thickness of 2 mm.}
    \label{fig:gamma-erg}
\end{figure}

The runaway generation rate is computed as
\begin{equation}
    \label{eq:compton-rate}
    \dot{n}_\mathrm{re} = n_\mathrm{e} \int
    \sigma_\mathrm{c}(E_\gamma)~\mathrm{d}[\phi_{\gamma,\mathrm{d}}(E_\gamma)]
\end{equation}
with $n_\mathrm{e}$ the nominal electron density and
$\phi_{\gamma,\mathrm{d}}(E_\gamma)$ the energy-dependent delayed gamma
radiation back-flux distribution. An example of this distribution is given in
\cref{fig:gamma-erg} and compared to the energy distribution for the total
(prompt plus
delayed) back-flux. The energy distribution of the delayed gamma radiation
back-flux is similar to that of the total back-flux, except for a lack of
high-energy gamma rays above a few MeV.

The energy-dependent Compton cross section, $\sigma_\mathrm{c}(E_\gamma)$, is
obtained from
\begin{equation}
    \label{eq:compton-xs}
    \sigma_\mathrm{c}(E_\gamma) = \int_{\theta_\mathrm{c}}^\pi
    \frac{\mathrm{d}\sigma}{\mathrm{d}\Omega}~\mathrm{d}\Omega
\end{equation}
where $\mathrm{d}\Omega = 2\pi\sin\theta~\mathrm{d}\theta$ and
\begin{equation}
    \label{eq:klein-nishina}
    \frac{\mathrm{d}\sigma}{\mathrm{d}\Omega} = \frac{r_\mathrm{e}^2}{2}
    \frac{{E_\gamma '}^2}{E_\gamma^2}\left(\frac{E_\gamma}{E_\gamma '} +
    \frac{E_\gamma '}{E_\gamma} - \sin^2\theta\right)
\end{equation}
with $E_\gamma '$ the energy of the scattered gamma ray and $\theta$ the
scattering angle. A critical value of the scattering angle exists,
$\theta_\mathrm{c}$, which depends on $E_\gamma$ and must be exceeded to
generate an electron with energy greater than $W_\mathrm{c}$. This is given by
\cite{Martin-Solis-2017-NuclFus}
\begin{equation}
    \label{eq:theta-crit}
    \theta_\mathrm{c} = \cos^{-1} \left(1 - \frac{m_e c^2}{E_\gamma}
    \frac{W_\mathrm{c} / E_\gamma}{1 - W_\mathrm{c} / E_\gamma}\right)
\end{equation}

The average velocity of a Compton scattering electron is obtained using
\cref{eq:klein-nishina} as
\begin{equation}
    \label{eq:electron-speed}
    \langle v_\mathrm{re}\rangle = \left\lbrace
    \int_{E_\gamma} \int_\Omega v_\mathrm{re}(\Omega)
    \frac{\mathrm{d}\sigma}{\mathrm{d}\Omega}
    ~\mathrm{d}\Omega~\mathrm{d}[\phi_{\gamma,\mathrm{d}}(E_\gamma)]
    \right\rbrace \left/\left\lbrace
    \int_{E_\gamma} \int_\Omega \frac{\mathrm{d}\sigma}{\mathrm{d}\Omega}
    ~\mathrm{d}\Omega~\mathrm{d}[\phi_{\gamma,\mathrm{d}}(E_\gamma)]
    \right\rbrace\right.
\end{equation}
where $v_\mathrm{re}(\Omega)$ of a Compton scattering electron may be computed
from its recoil energy, $E_\gamma - E_\gamma '$.

Finally, we estimate the value of $f(\xi_\mathrm{c})$ by assuming an isotropic
scattering distribution, due to the fact that gamma radiation back-flux is
emitted from every plasma-facing surface over a range of exit angles. Thus, we
use the simple relation $f(\xi_\mathrm{c}) = \cos^{-1}(\xi_\mathrm{c}) / \pi$.

\begin{figure}[htb]
    \centering
    \includegraphics[width=0.475\linewidth]{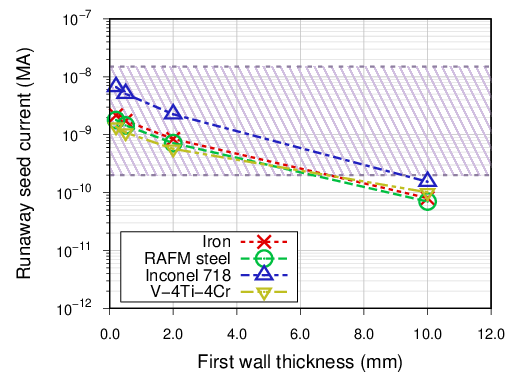}
    \caption{Compton scattering runaway electron seed currents,
        $I_\mathrm{seed}$ (in MA), estimated from MCNP calculations of delayed
        gamma radiation back-fluxes for various material configurations. The
        shaded region depicts the range of Compton seed current values
        calculated in Fig. 9 of \cite{Martin-Solis-2017-NuclFus}.}
    \label{fig:compton}
\end{figure}

\cref{fig:compton} gives the results of these calculations ($I_\mathrm{seed}$)
for each combination of SM choice and FW thickness simulated for this work.
For these estimates, we choose an ITER-like geometry ($R_0 = 6.2~\text{m}$, $a
= 2.0~\text{m}$, plasma chamber volume $V = 840~\text{m}^3$, and wall surface
area $A_\mathrm{w} = 660~\text{m}^2$), yielding parameter values of
$A_\mathrm{xs} = V / (2\pi R_0) = 21.6~\text{m}^2$, $\xi_\mathrm{c} = 0.803$,
and $f(\xi_\mathrm{c}) = 0.203$. We choose $W_\mathrm{c} = 18.6~\text{keV}$,
which is the minimum value of $W_\mathrm{c}$ required to eliminate the
possibility of runaway seeding by tritium decay. We also choose
$\tau_\mathrm{res} = 0.034~\text{s}$ to allow comparison with some of the
results from \cite{Martin-Solis-2017-NuclFus}. We note that the use of
$\tau_\mathrm{res}$ in \cref{eq:compton-rate} is an approximation, necessary
in the absence of plasma physics simulations, by which we estimate the Compton
scattering runaway seed current generated on the same time scale as the
current quench decay time. Finally, we choose representative values of the
fusion neutron flux to the wall, $\phi_\mathrm{n} \sim 2.68\times
10^{17}~\text{m}^{-2}\text{s}^{-1}$, representing 500 MW of fusion power, and
nominal electron density during the disruption of $n_\mathrm{e} = 4\times
10^{20}~\text{m}^{-3}$ which includes the effect of impurity injection
\cite{Martin-Solis-2015-PhysPlas}. For this analysis, we use the terminal
delayed gamma radiation back-flux levels (as reported in \cref{tab:summary}).

The runaway seed current data in \cref{fig:compton} highlight the impact of
material configuration parameters, as both the FW thickness and SM choice
significantly affect the seed current. A sufficient first wall thickness can
greatly attenuate the gamma radiation back-flux, particularly the lower-energy
gamma rays which have higher Compton scattering cross section values, and thus
reduce the runaway seed current accordingly. In particular, increasing the
first wall thickness from 2 mm to 10 mm reduces the runaway seed current by a
factor of 5 to 10. The choice of structural material has a similarly large
effect, most notably the runaway seed current when Inconel 718 is the
structural material is consistently larger than that in other cases by a
factor of 2 to 5. We also note that different SM choices and FW thicknesses
interact to produce a complex variation of $I_\mathrm{seed}$. For example,
V-4Ti-4Cr SM exhibits the lowest runaway seed currents for FW thicknesses of 2
mm or less, but when the FW thickness is 10 mm the iron-based SMs exhibit the
lowest seed currents.

We note that the values of $I_\mathrm{seed}$ in \cref{fig:compton} agree well
with previous published studies of this topic. Mart{\'i}n-Sol{\'i}s and
coworkers calculated Compton runaway seed values ranging from $2\times
10^{-10}$ to $1.5\times 10^{-8}~\text{MA}$ \cite{Martin-Solis-2017-NuclFus}.
Vallhagen and coworkers performed a similar analysis \cite{Vallhagen-2020-JPP,
    Vallhagen-2024-NF}, and reported total runaway seed currents from all
mechanisms in the range from $10^{-12}$ to $10^{-6}~\text{MA}$ under a variety
of plasma conditions (although Compton-only contributions were not separated
in these works). These authors showed that even these apparently small
magnitudes of runaway seed current are sufficient to avalanche up to $\sim$MA
runaway currents which can threaten to damage or destroy reactor components.
However, this work used an unrealistically large value of $\phi_\gamma \sim
10^{18}~\text{m}^{-2}\text{s}^{-1}$ and did not reduce this value to account
for the loss of prompt back-flux after the thermal quench.

Comparing the results in \cref{fig:compton} to these literature values
indicates the value of the present work. On one hand, the good agreement
between our estimates and the wider literature is encouraging, given the only
cursory consideration of plasma conditions possible in our estimate. At the
same time, we note that the previous calculations were performed using a rough
estimate of the gamma radiation back-flux energy spectrum, which is of
questionable accuracy (notably, the high-energy tail where $E_\gamma >
10~\text{MeV}$ is unphysical) and was computed for a Be first wall but never
updated for the planned tungsten first wall. The ready availability of
accurate gamma radiation back-flux magnitudes and energy spectra via the
simulation approach shown in this work, which can be readily prepared for
specific material configurations, enables accurate predictions of the Compton
seed current. Therefore, a coupled material-plasma modeling framework to make
these calculations on-demand is a key direction for future work.

\subsection{Electron back-flux impact on sheath dynamics}
\label{subsec:sheath-backflux}

Finally, electron back-fluxes of $\sim$MeV average energies can impact the
plasma sheath in several ways, depending on the angular distribution of the
emitted electrons and the magnetic incident angle at the wall. Firstly, with a
nearly tangential incident angle for the magnetic field at the divertor and
first wall, one would expect that the emitted electrons would follow the
gyro-orbit and impact the wall surface again. Both high energy backscattered
electrons (BSE) and the predominantly low energy secondary emission electrons
(SEE) can result from this re-entry of the gyrating electrons into the wall,
which could qualitatively modify the sheath behavior from classical models
\cite{Campanell-2013-PhysRevE, Campanell-2016-PhysRevLett,
    Hobbs-1967-PlasPhys, Bradshaw-2024-PSST}. We note that the gyroradius of
    the
high-energy electrons is much larger than that of the low temperature ions in
the divertor, so that the Chodura sheath physics \cite{Chodura-1982-PhysFluid,
    Stangeby-2012-NuclFus} should be revisited. Secondly, the high-energy
electrons may affect the collisional-radiative physics due to the relativistic
enhancement of the collisional excitation and ionization cross sections
\cite{Garland-2020-PhysPlas}, which can mostly occur near the target surface
where the high-energy electrons get trapped. Finally, if the emitted high
energy electrons were able slide away along the magnetic field before gyrating
back to the wall, their high energies allow a large electron current from the
wall emission, which can alter or potentially even reverse the sheath
potential \cite{Campanell-2013-PhysRevE, Campanell-2016-PhysRevLett} despite a
lower back-flux magnitude.

\begin{figure}[htb]
    \centering
    \begin{subfigure}{0.475\linewidth}
        \centering
        \includegraphics[width=\linewidth]{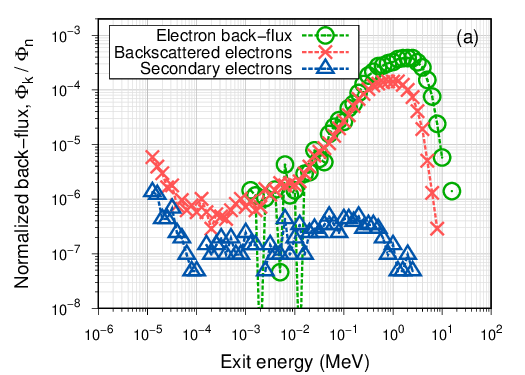}
        \phantomcaption
        \label{subfig:see-erg}
    \end{subfigure}
    \begin{subfigure}{0.475\linewidth}
        \centering
        \includegraphics[width=\linewidth]{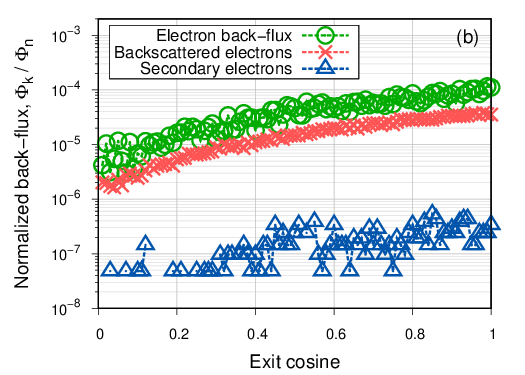}
        \phantomcaption
        \label{subfig:see-cos}
    \end{subfigure}
    \caption{(\subref{subfig:see-erg}) Energy and (\subref{subfig:see-cos})
        cosine distributions for electron back-flux (from incident neutrons),
        backscattered electrons, and secondary electrons (both from backflux
        rebombardment) emitted from RAFM steel with a 2-mm first wall
        thickness.
        Note that the BSE and SEE distributions obtained from MCNP calculations
        are necessarily incomplete, owing to the 1 keV lower cutoff energy.}
    \label{fig:see-dists}
\end{figure}

While analysis of these potential impacts requires detailed plasma
simulations and is beyond the scope of the current work, the energy and
angular distributions of the BSEs and SEEs obtained from MCNP calculations are
useful results which form the input conditions for plasma simulations.
\cref{fig:see-dists} shows such BSE and SEE energy and cosine distributions
exiting the surface obtained by MCNP calculations for RAFM SM with a 2-mm FW
thickness, compared against the electron back-flux (from neutron irradiation)
for the same material configuration. The simulation setup is same as Fig.
\ref{fig:mcnp-setup} except that the normal incident particles are electrons
with the same energy distribution as the electron back-flux, e.g., as in
\cref{subfig:erg-rafm2} We must note that the BSE and SEE distributions
obtained from MCNP calculations are necessarily incomplete, owing to the
absolute lower energy cutoff for electrons in MCNP of 10 eV. The results show
that above 10 eV, the backscattering yield is 0.412 BSEs per incident
electron, while the secondary emission yield is much smaller at only 0.0034
SEEs per incident electron. In particular, the missing low-energy part of the
SEE distribution is likely to contribute significantly to the total SEE
fluxes, given that the energy distributions for SEEs increases sharply to a
maximum as the energy approaches the 10-eV lower cutoff. Therefore, for
complete SEE input to plasma sheath simulations, MCNP calculations must be
supplemented by low-energy SEE models \cite{Bradshaw-2024-PSST}.

%
\section{Conclusion}
\label{sec:conclusion}

We have quantified the radiation back-fluxes from fusion neutron irradiation
of reactor materials with MCNP simulations. These radiation back-fluxes are
remarkably large, with neutron and gamma ray back-fluxes on the same
magnitude as the incident fusion neutron flux. Though strongly correlated with
the gamma ray back-fluxes, the electron back-fluxes are two
orders of magnitude smaller in all cases. The
radiation back-fluxes are significantly impacted by the material design
parameters. The neutron back-flux depends most strongly on the choice of
structural material, while the gamma ray and electron back-flux magnitudes are
dominated by the first wall thickness. In addition to prompt back-fluxes,
nuclear decay processes in activated reactor materials lead to significant
delayed gamma ray and electron back-fluxes. During steady-state fusion power
operations, the delayed gamma ray back-flux builds up to a magnitude of
2\%--7\% of the prompt back-flux magnitude, while the delayed electron
back-flux magnitude remains two orders of magnitude lower.

Radiation back-fluxes can have profound impacts on plasma performance and
reactor operations. The significant neutron back-flux has the ultimate effect
of increasing the gamma radiation and electron emission back-fluxes by
$\sim$50--80\%, although in a real reactor this enhancement also depends on
the geometry of the plasma-facing surfaces. The substantial neutron back-flux
also implies significant neutron multiplication in the first wall and
structural materials, which must be considered in the design of tritium
breeding blankets and could ease the design constraints of those systems.
Furthermore, the build-up over time of delayed gamma radiation back-fluxes
during fusion power operations will induce significant runaway electron
seeding by Compton scattering of cold plasma electrons, posing significant
challenges for disruption mitigation compared to non-nuclear environments.
Finally, the electron back-flux of $\sim$MeV average energies can impact the
plasma sheath in several ways: electrons may be directed back into the surface
by the magnetic field, leading to significant backscattering and secondary
electron emission fluxes into the sheath, they may drive enhanced excitation
and ionization of near the wall, or they may slide away from the wall
along the magnetic field, contributing a large electron emission current
component which could drastically alter the sheath potential. All of these
back-flux phenomena can significantly impact plasma performance in a future
fusion power reactor.

We envision two principal avenues for future extension and implementation of
this work. First, our simulation methodology should be scaled up to reactor
geometries to obtain spatially-dependent radiation back-flux profiles for real
devices, including studies of potential TBB designs. Second, the radiation
back-flux profiles must be coupled to whole-device plasma physics simulations
to elucidate the impact on plasma performance. Such efforts are urgently
motivated by the revelation from this work of how impactful the effects of
radiation back-fluxes can be. Given the significant potential impact of these
formerly overlooked mechanisms on the performance of fusion power reactors,
analysis of radiation back-flux effects are of crucial importance to any
consideration of reactor engineering and design.

%
\section*{Acknowledgements}

This work is funded by the U.S. Department of Energy Office of Fusion Energy
Sciences (DOE-FES) under the Tokamak Disruption Simulation (TDS) Scientific
Discovery through Advanced Computing (SciDAC) project at Los Alamos National
Laboratory (LANL) under Contract No. 89233218CNA000001. Los Alamos National
Laboratory is operated by Triad National Security LLC, for the National
Nuclear Security administration of the U.S. DOE under Contract No.
89233218CNA0000001.

The authors are grateful to Colin J. Josey for helpful discussion regarding
the \texttt{ACT} card in MCNP.

\paragraph{CRediT author statement} \textbf{Michael A. Lively:}
Conceptualization, Methodology, Formal Analysis, Investigation, Writing -
original draft, Writing - review \& editing, Visualization. \textbf{Danny
    Perez:} Conceptualization, Writing - review \& editing. \textbf{Blas P.
    Uberuaga:} Conceptualization, Writing - review \& editing. \textbf{Yanzeng
    Zhang:} Conceptualization, Writing - review \& editing. \textbf{Xianzhu
    Tang:}
Conceptualization, Writing - review \& editing, Project administration,
Funding acquisition.

%
\bibliographystyle{elsarticle-num}
\bibliography{backflux-refs}

\begin{thebibliography}{10}
\expandafter\ifx\csname url\endcsname\relax
  \def\url#1{\texttt{#1}}\fi
\expandafter\ifx\csname urlprefix\endcsname\relax\def\urlprefix{URL }\fi
\expandafter\ifx\csname href\endcsname\relax
  \def\href#1#2{#2} \def\path#1{#1}\fi

\bibitem{ITER-RP-2024}
D.~J. Campbell, A.~Loarte, D.~Boilson, X.~Bonnin, P.~de~Vries, L.~Giancarli,
  Y.~Gribov, M.~Henderson, S.~H. Kim, P.~Lamalle, M.~Lehnen, T.~Luce, I.~Nunes,
  A.~R. Polevoi, S.~D. Pinches, R.~A. Pitts, R.~Reichle, M.~Schneider,
  J.~Snipes, J.~van~der Laan, G.~Vayakis, {ITER IRP Contributors},
  \href{https://www.iter.org/technical-reports?id=26}{{ITER} research plan
  within the staged approach ({Level III} - final version)}, Tech. Rep.
  ITR-24-005, ITER Organization, St Paul-lez-Durance, France (April 2024).
\newline\urlprefix\url{https://www.iter.org/technical-reports?id=26}

\bibitem{Ishino-1996-JNuclMater}
S.~Ishino, Implicatons of fundamental radiation damage studies in the research
  and development of materials for a fusion reactor, Journal of Nuclear
  Materials 239 (1996) 24--33.
\newblock \href {https://doi.org/10.1016/S0022-3115(96)00486-2}
  {\path{doi:10.1016/S0022-3115(96)00486-2}}.

\bibitem{Mata-2011-FusEngDes}
F.~Mata, R.~Vila, C.~Ortiz, N.~Casal, A.~Ibarra, D.~Rapisarda, V.~Queral,
  Analysis of displacement damage in materials in nuclear fusion facilities
  ({DEMO}, {IMFIF}, and {TechnoFusion}), Fusion Engineering and Design
  86~(9-11) (2011) 2425--2428.
\newblock \href {https://doi.org/10.1016/j.fusengdes.2010.12.041}
  {\path{doi:10.1016/j.fusengdes.2010.12.041}}.

\bibitem{Rubel-2019-JFusEne}
M.~Rubel, Fusion neutrons: Tritium breeding and impact on wall materials and
  components of diagnostic systems, Journal of Fusion Energy 38 (2019)
  315--329.
\newblock \href {https://doi.org/10.1007/s10894-018-0182-1}
  {\path{doi:10.1007/s10894-018-0182-1}}.

\bibitem{Spitsyn-2019-FusEngDes}
A.~V. Spitsyn, N.~P. Bobyr, T.~V. Kulevoy, P.~A. Fedin, A.~I. Semennikov, V.~S.
  Stolbunov, Use of {MeV} energy ion accelerators to simulate the neutron
  damage in fusion reactor materials, Fusion Engineering and Design 146~(A)
  (2015) 1313--1316.
\newblock \href {https://doi.org/10.1016/j.fusengdes.2019.02.065}
  {\path{doi:10.1016/j.fusengdes.2019.02.065}}.

\bibitem{Breidokaite-2023-RadPhysChem}
S.~Breidokaite, G.~Stankunas, Helium production and material damage rate
  assessment in {EU DEMO HCPB} divertor, Radiation Physics and Chemistry 210
  (2023) 111024.
\newblock \href {https://doi.org/10.1016/j.radphyschem.2023.111024}
  {\path{doi:10.1016/j.radphyschem.2023.111024}}.

\bibitem{Li-2024-JNuclMater}
L.~Li, Y.~Hu, L.~Peng, J.~Shi, Y.~Sun, X.~Hu, C.~Hu, Helium bubble evolution
  under cascade in bcc iron relevant to fusion conditions investigated by a
  novel coupling {MD-OKMC} method, Journal of Nuclear Materials 591 (2024)
  154908.
\newblock \href {https://doi.org/10.1016/j.jnucmat.2024.154908}
  {\path{doi:10.1016/j.jnucmat.2024.154908}}.

\bibitem{Hernandez-2018-FusEngDes}
F.~A. Hern{\'a}ndez, P.~Pereslavtsev, First principles review of options for
  tritium breeder and neutron multiplier materials for breeding blankets in
  fusion reactors, Fusion Engineering and Design 137 (2018) 243--256.
\newblock \href {https://doi.org/10.1016/j.fusengdes.2018.09.014}
  {\path{doi:10.1016/j.fusengdes.2018.09.014}}.

\bibitem{Segantin-2020-FusEngDes}
S.~Segantin, R.~Testoni, Z.~Hartwig, D.~Whyte, M.~Zucchetti, Optimization of
  tritium breeding ratio in {ARC} reactor, Fusion Engineering and Design 154
  (2020) 111531.
\newblock \href {https://doi.org/10.1016/j.fusengdes.2020.111531}
  {\path{doi:10.1016/j.fusengdes.2020.111531}}.

\bibitem{Bae-2024-NuclFus}
J.~W. Bae, D.~Young, K.~Borowiec, V.~Badalassi, Integral analysis of the effect
  of material dimension and composition on tokamak neutronics, Nuclear Fusion
  64~(5) (2024) 056013.
\newblock \href {https://doi.org/10.1088/1741-4326/ad33ee}
  {\path{doi:10.1088/1741-4326/ad33ee}}.

\bibitem{Kim-2024-FusEngDes}
B.~C. Kim, Exploratory neutronic evaluation on the enhancement of tritium
  breeding and energy multiplication capability using
  {(Th,U)O\textsubscript{2}} in a solid-state tritium breeding blanket of a
  fusion demonstration reactor, Fusion Engineering and Design 205 (2024)
  114558.
\newblock \href {https://doi.org/10.1016/j.fusengdes.2024.114558}
  {\path{doi:10.1016/j.fusengdes.2024.114558}}.

\bibitem{Prost-2024-JNuclMater}
V.~Prost, S.~Ogier-Collin, F.~A. Volpe, Compact fusion blanket using plasma
  facing liquid {Li-LiH} walls and {Pb} pebbles, Journal of Nuclear Materials
  599 (2024) 155239.
\newblock \href {https://doi.org/10.1016/j.jnucmat.2024.155239}
  {\path{doi:10.1016/j.jnucmat.2024.155239}}.

\bibitem{Ghani-2015-FusEngDes}
Z.~Ghani, A.~Turner, S.~Mangham, J.~Naish, M.~Lis, L.~Packer, M.~Laughlin,
  Radiation levels in the {ITER} tokamak complex during and after plasma
  operation, Fusion Engineering and Design 96-97 (2015) 261--264.
\newblock \href {https://doi.org/10.1016/j.fusengdes.2015.05.019}
  {\path{doi:10.1016/j.fusengdes.2015.05.019}}.

\bibitem{Wilson-2018-FusSciTech}
S.~C. Wilson, S.~W. Mosher, K.~E. Royston, C.~R. Daily, A.~M. Ibrahim,
  Validation of the {MS-CADIS} method for full-scale shutdown dose rate
  analysis, Fusion Science and Technology 74~(4) (2018) 288--302.
\newblock \href {https://doi.org/10.1080/15361055.2018.1483687}
  {\path{doi:10.1080/15361055.2018.1483687}}.

\bibitem{Royston-2019-FusSciTech}
K.~Royston, G.~Radulescu, W.~van Hove, S.~Wilson, S.~Kim, Assessment of
  activation on level {L3} of the tokamak building due to the {ITER} tokamak
  cooling water system, Fusion Science and Technology 75~(6) (2019) 458--465.
\newblock \href {https://doi.org/10.1080/15361055.2019.1606519}
  {\path{doi:10.1080/15361055.2019.1606519}}.

\bibitem{Qiu-2024-FusEngDes}
Y.~Qiu, M.~Ansorge, I.~{\'A}lvarez, K.~Ambro{\v z}i{\v c}, T.~Berry,
  B.~Bie{\'n}kowska, H.~Chohan, A.~{\v C}ufar, D.~Dworak, T.~Dezsi, T.~Eade,
  J.~Garc{\'i}a, D.~Jimenez-Rey, I.~Lengar, A.~Lopez-Revelles, V.~Lopez,
  E.~Mendoza, F.~Mota, M.~Martinez-Echevarria, F.~Ogando, J.~Park,
  T.~Piotrowski, A.~Serikov, G.~Stankunas, A.~Tidikas, G.~Tracz, G.~{\v
  Z}erovnik, F.~Arbeiter, F.~Arranz, S.~Becerril, P.~Cara, D.~Bernardi,
  J.~Castellanos, J.~Guti{\'e}rrez, A.~Ibarra, W.~Kr{\'o}las, J.~Maestre,
  F.~Mart{\'i}n-Fuertes, J.~Marug{\'a}n, G.~Miccich{\'e},
  J.~Mart{\'i}nez-Serrano, F.~Nitti, I.~Podadera, U.~Wi{\c a}cek, U.~Fischer,
  Overview of recent advancement in {IFMIF-DONES} neutronics activities, Fusion
  Engineering and Design 201 (2024) 114242.
\newblock \href {https://doi.org/10.1016/j.fusengdes.2024.114242}
  {\path{doi:10.1016/j.fusengdes.2024.114242}}.

\bibitem{El-Guebaly-2005-FusSciTech}
L.~El-Guebaly, P.~Wilson, D.~Paige, {ARIES Team}, Initial activation assessment
  for {ARIES} compact stellarator power plant, Fusion Science and Technology
  47~(3) (2005) 440--444.
\newblock \href {https://doi.org/10.13182/FST05-A726}
  {\path{doi:10.13182/FST05-A726}}.

\bibitem{Pampin-2012-FusEngDes}
R.~Pampin, S.~Zheng, S.~Lilley, B.~C. Na, M.~J. Laughlin, N.~P. Taylor,
  Activation analyses updating the {ITER} radioactive waste assessment, Fusion
  Engineering and Design 87~(7-8) (2012) 1230--1234.
\newblock \href {https://doi.org/10.1016/j.fusengdes.2012.02.110}
  {\path{doi:10.1016/j.fusengdes.2012.02.110}}.

\bibitem{Someya-2017-FusSciTech}
Y.~Someya, K.~Tobita, H.~Utoh, N.~Asakura, Y.~Sakamoto, K.~Hoshino,
  M.~Nakamura, S.~Tokunaga, Management strategy for radioactive waste in the
  fusion {DEMO} reactor, Fusion Science and Technology 68~(2) (2017) 423--427.
\newblock \href {https://doi.org/10.1016/j.fusengdes.2012.02.110}
  {\path{doi:10.1016/j.fusengdes.2012.02.110}}.

\bibitem{Gilbert-2018-FusEngDes}
M.~R. Gilbert, T.~Eade, C.~Bachmann, U.~Fischer, N.~P. Taylor, Waste assessment
  of {European DEMO} fusion reactor designs, Fusion Engineering and Design
  136~(A) (2018) 42--48.
\newblock \href {https://doi.org/10.1016/j.fusengdes.2017.12.019}
  {\path{doi:10.1016/j.fusengdes.2017.12.019}}.

\bibitem{Cao-2021-FusEngDes}
Q.~Cao, X.~Wang, M.~Yin, S.~Qu, L.~Zhang, B.~Zhou, F.~Zhao, Preliminary
  activation analysis and radioactive waste classification for {CFETR}, Fusion
  Engineering and Design 172 (2021) 112789.
\newblock \href {https://doi.org/10.1016/j.fusengdes.2021.112789}
  {\path{doi:10.1016/j.fusengdes.2021.112789}}.

\bibitem{Martin-Solis-2017-NuclFus}
J.~Martín-Solís, A.~Loarte, M.~Lehnen, Formation and termination of runaway
  beams in {ITER} disruptions, Nuclear Fusion 57~(6) (2017) 066025.
\newblock \href {https://doi.org/10.1088/1741-4326/aa6939}
  {\path{doi:10.1088/1741-4326/aa6939}}.

\bibitem{Vallhagen-2020-JPP}
O.~Vallhagen, O.~Embreus, I.~Pusztai, L.~Hesslow, T.~F{\"u}l{\"o}p, Runaway
  dynamics in the {DT} phase of {ITER} operations in the presence of massive
  material injection, Journal of Plasma Physics 86~(4) (2020) 475860401.
\newblock \href {https://doi.org/10.1017/S0022377820000859}
  {\path{doi:10.1017/S0022377820000859}}.

\bibitem{Vallhagen-2024-NF}
O.~Vallhagen, L.~Hanebring, F.~J. Artola, M.~Lehnen, E.~Nardon,
  T.~F{\"u}l{\"o}p, M.~Hoppe, S.~L. Newton, I.~Pusztai, Runaway electron
  dynamics in {ITER} disruptions with shattered pellet injection, Nuclear
  Fusion 64~(8) (2024) 086033.
\newblock \href {https://doi.org/10.1088/1741-4326/ad54d7}
  {\path{doi:10.1088/1741-4326/ad54d7}}.

\bibitem{Campanell-2013-PhysRevE}
M.~D. Campanell, Negative plasma potential relative to electron-emitting
  surfaces, Physical Review E 88~(3) (2013) 033103.
\newblock \href {https://doi.org/10.1103/PhysRevE.88.033103}
  {\path{doi:10.1103/PhysRevE.88.033103}}.

\bibitem{Campanell-2016-PhysRevLett}
M.~D. Campanell, M.~V. Umansky, Strongly emitting surfaces unable to float
  below plasma potential, Physical Review Letters 116~(8) (2016) 085003.
\newblock \href {https://doi.org/10.1103/PhysRevLett.116.085003}
  {\path{doi:10.1103/PhysRevLett.116.085003}}.

\bibitem{Bradshaw-2024-PSST}
K.~Bradshaw, B.~Srinivasan, Energy-dependent implementation of secondary
  electron emission models in continuum kinetic sheath simulations, Plasma
  Sources Science and Technology 33~(3) (2024) 035008.

\bibitem{MCNP6-3}
J.~A. Kulesza, T.~R. Adams, J.~C. Armstrong, S.~R. Bolding, F.~B. Brown, J.~S.
  Bull, T.~P. Burke, A.~R. Clark, R.~A. Forster, III, J.~F. Giron, T.~S.
  Grieve, C.~J. Josey, R.~L. Martz, G.~W. McKinney, E.~J. Pearson, M.~E.
  Rising, C.~J. Solomon, Jr., S.~Swaminarayan, T.~J. Trahan, S.~C. Wilson,
  A.~J. Zukaitis, {MCNP\textsuperscript{\textregistered}} code version 6.3.0
  theory \& user manual, Tech. Rep. LA-UR-22-30006, Los Alamos National
  Laboratory, Los Alamos, NM, USA (2022).
\newblock \href {https://doi.org/10.2172/1889957} {\path{doi:10.2172/1889957}}.

\bibitem{Kohyama-1996-JNuclMater}
A.~Kohyama, A.~Hishinuma, D.~S.~G. andR. L.~Klueh, W.~Dietz, K.~Ehrlich,
  Low-activation ferritic and martensitic steels for fusion application,
  Journal of Nuclear Materials 233-237 (1996) 138--147.
\newblock \href {https://doi.org/10.1016/S0022-3115(96)00327-3}
  {\path{doi:10.1016/S0022-3115(96)00327-3}}.

\bibitem{Federici-2017-NuclFus}
G.~Federici, W.~Biel, M.~R. Gilbert, R.~Kemp, N.~Taylor, R.~Wenninger,
  {European DEMO} design strategy and consequences for materials, Nuclear
  Fusion 57~(9) (2017) 092002.
\newblock \href {https://doi.org/10.1088/1741-4326/57/9/092002}
  {\path{doi:10.1088/1741-4326/57/9/092002}}.

\bibitem{Bae-2022-NuclFus}
J.~W. Bae, E.~peterson, J.~Shimwell, {ARC} reactor neutronics multi-code
  validation, Nuclear Fusion 62~(6) (2022) 066016.
\newblock \href {https://doi.org/10.1088/1741-4326/ac5450}
  {\path{doi:10.1088/1741-4326/ac5450}}.

\bibitem{Kurtz-2004-JNuclMater}
R.~J. Kurtz, K.~Abe, V.~M. Chernov, D.~T. Hoelzer, H.~Matsui, T.~Muroga, G.~R.
  Odette, Recent progress on development of vanadium alloys for fusion, Journal
  of Nuclear Materials 329-333 (2004) 47--55.
\newblock \href {https://doi.org/doi:10.1016/j.jnucmat.2004.04.299}
  {\path{doi:doi:10.1016/j.jnucmat.2004.04.299}}.

\bibitem{Brown-2018-NDS}
D.~Brown, M.~Chadwick, R.~Capote, A.~Kahler, A.~Trkov, M.~Herman, A.~Sonzogni,
  Y.~Danon, A.~Carlson, M.~Dunn, D.~Smith, G.~Hale, G.~Arbanas, R.~Arcilla,
  C.~Bates, B.~Beck, B.~Becker, F.~Brown, R.~Casperson, J.~Conlin, D.~Cullen,
  M.-A. Descalle, R.~Firestone, T.~Gaines, K.~Guber, A.~Hawari, J.~Holmes,
  T.~Johnson, T.~Kawano, B.~Kiedrowski, A.~Koning, S.~Kopecky, L.~Leal,
  J.~Lestone, C.~Lubitz, J.~M. Dami{\'a}n, C.~Mattoon, E.~McCutchan,
  S.~Mughabghab, P.~Navratil, D.~Neudecker, G.~Nobre, G.~Noguere, M.~Paris,
  M.~Pigni, A.~Plompen, B.~Pritychenko, V.~Pronyaev, D.~Roubtsov, D.~Rochman,
  P.~Romano, P.~Schillebeeckx, S.~Simakov, M.~Sin, I.~Sirakov, B.~Sleaford,
  V.~Sobes, E.~Soukhovitskii, I.~Stetcu, P.~Talou, I.~Thompson, S.~van~der
  Marck, L.~Welser-Sherrill, D.~Wiarda, M.~White, J.~Wormald, R.~Wright,
  M.~Zerkle, G.~\v{Z}erovnik, Y.~Zhu, {ENDF/B-VIII.0}: The 8th major release of
  the nuclear data library with cielo-project cross sections, new standards,
  and thermal scattering data, Nuclear Data Sheets 148 (2018) 1--142.
\newblock \href {https://doi.org/10.1016/j.nds.2018.02.001}
  {\path{doi:10.1016/j.nds.2018.02.001}}.

\bibitem{Conlin-2018-LAUR}
J.~L. Conlin, W.~Haeck, D.~Neudecker, D.~K. Parsons, M.~C. White, Release of
  {ENDF/B-VIII.0}-based {ACE} data files, Tech. Rep. LA-UR-18-24034, Los Alamos
  National Laboratory, Los Alamos, NM, United States (May 2018).
\newblock \href {https://doi.org/10.2172/1438139} {\path{doi:10.2172/1438139}}.

\bibitem{Cullen-2014-EPICS}
D.~E. Cullen, {EPICS2014}: Electron photon interaction cross sections (version
  2014), Tech. Rep. IAEA-NDS-218, International Atomic Energy Agency - Nuclear
  Data Section, Vienna, Austria (2014).

\bibitem{Hughes-2014-PNST}
H.~G. Hughes, Recent developments in low-energy electron/photon transport for
  {MCNP6}, Progress in Nuclear Science and Technology 4 (2014) 454--458.
\newblock \href {https://doi.org/10.15669/pnst.4.454}
  {\path{doi:10.15669/pnst.4.454}}.

\bibitem{Hughes-2017-ICRS}
H.~G. Hughes, Improvements in electron-photon-relaxation data for {MCNP6}, EPJ
  Web of Conferences 153 (2017) 06009.
\newblock \href {https://doi.org/10.1051/epjconf/201715306009}
  {\path{doi:10.1051/epjconf/201715306009}}.

\bibitem{Berger-1988-MCTrans}
M.~J. Berger, Electron stopping powers for transport calculations, in:
  T.~Jenkins, W.~Nelson, A.~Rindi (Eds.), {Monte Carlo} Transport of Electrons
  and Photons, Plenum Press, New York, NY, 10013, United States of America,
  1988, Ch.~3, pp. 57--80.
\newblock \href {https://doi.org/10.1007/978-1-4613-1059-4}
  {\path{doi:10.1007/978-1-4613-1059-4}}.

\bibitem{Seltzer-1988-MCTrans}
S.~M. Seltzer, Cross sections for bremstrahhlung production and electron-impact
  ionization, in: T.~M. Jenkins, W.~R. Nelson, A.~Rindi (Eds.), {Monte Carlo}
  Transport of Electrons and Photons, Plenum Press, New York, NY, 10013, United
  States of America, 1988, Ch.~4, pp. 81--114.
\newblock \href {https://doi.org/10.1007/978-1-4613-1059-4}
  {\path{doi:10.1007/978-1-4613-1059-4}}.

\bibitem{Adams-2000-LAUR}
K.~J. Adams, Electron upgrade for {MCNP4B}, Tech. Rep. LA-UR-00-3581, Los
  Alamos National Laboratory, Los Alamos, NM, United States (2000).

\bibitem{Chadwick-2006-NDS}
M.~Chadwick, P.~Oblo{\v z}insk{\' y}, M.~Herman, N.~Greene, R.~McKnight,
  D.~Smith, P.~Young, R.~MacFarlane, G.~Hale, S.~Frankle, A.~Kahler, T.~Kawano,
  R.~Little, D.~Madland, P.~Moller, R.~Mosteller, P.~Page, P.~Talou,
  H.~Trellue, M.~White, W.~Wilson, R.~Arcilla, C.~Dunford, S.~Mughabghab,
  B.~Pritychenko, D.~Rochman, A.~Sonzogni, C.~Lubitz, T.~Trumbull, J.~Weinman,
  D.~Brown, D.~Cullen, D.~Heinrichs, D.~McNabb, H.~Derrien, M.~Dunn, N.~Larson,
  L.~Leal, A.~Carlson, R.~Block, J.~Briggs, E.~Cheng, H.~Huria, M.~Zerkle,
  K.~Kozier, A.~Courcelle, V.~Pronyaev, S.~van~der Marck, {ENDF/B-VII.0}: Next
  generation evaluated nuclear data library for nuclear science and technology,
  Nuclear Data Sheets 107~(12) (2006) 2931--3118.
\newblock \href {https://doi.org/10.1016/j.nds.2006.11.001}
  {\path{doi:10.1016/j.nds.2006.11.001}}.

\bibitem{Parsons-2022-LAUR}
D.~K. Parsons, Verification of the {ENDF7U} photonuclear data library for
  {MCNP}, Tech. Rep. LA-UR-22-25692, Los Alamos National Laboratory, Los
  Alamos, NM, United States (June 2022).

\bibitem{CINDER-90}
W.~B. Wilson, S.~T. Cowell, T.~R. England, A.~C. Hayes, P.~Moller, A manual for
  {CINDER'90} verion 07.4 codes and data, Tech. Rep. LA-UR-07-8412, Los Alamos
  National Laboratory, Los Alamos, NM, United States (March 2008).

\bibitem{Josey-2024-email}
C.~J. Josey, private communication (April 2024).

\bibitem{Kawamura-2003-NuclFus}
H.~Kawamura, E.~Ishitsuka, K.~Tsuchiya, M.~Nakamichi, M.~Uchida, H.~Yamada,
  K.~Nakamura, H.~Ito, T.~Nakazawa, H.~Takahashi, S.~Tanaka, N.~Y.~S. Kato,
  Y.~Ito, Development of advanced blanket materials for a solid breeder blanket
  of a fusion reactor, Nuclear Fusion 43~(8) (2003) 675--680.
\newblock \href {https://doi.org/10.1088/0029-5515/43/8/306}
  {\path{doi:10.1088/0029-5515/43/8/306}}.

\bibitem{Boccaccini-2022-FusEngDes}
L.~V. Boccaccini, F.~Arbeiter, P.~Arena, J.~Aubert, L.~B{\"u}hler,
  I.~Cristescu, A.~{Del Novo}, M.~Eboli, L.~Forest, C.~Harrington,
  F.~Hernandez, R.~Knitter, H.~Neuberger, D.~Rapisarda, P.~Sardain, G.~A.
  Spagnuolo, M.~Utili, L.~Vala, A.~Venturini, P.~Vladimirov, G.~Zhou, Status of
  maturation of critical technologies and systems design: Breeding blanket,
  Fusion Engineering and Design 179 (2022) 113116.
\newblock \href {https://doi.org/0.1016/j.fusengdes.2022.113116}
  {\path{doi:0.1016/j.fusengdes.2022.113116}}.

\bibitem{Mukai-2023-NuclMatEne}
K.~Mukai, J.-H. Kim, M.~Nakamichi, Measurement of thermal expansion anisotropy
  in be\textsubscript{12}ti and be\textsubscript{12}v, Nuclear Materials and
  Energy 36 (2023) 101473.
\newblock \href {https://doi.org/10.1016/j.nme.2023.101473}
  {\path{doi:10.1016/j.nme.2023.101473}}.

\bibitem{Zhou-2023-Energies}
G.~Zhou, F.~A. Hern{\'a}ndez, P.~Pereslavtsev, B.~Kiss, A.~Retheesh,
  L.~Maqueda, J.~H. Park, The {European DEMO} helium cooled pebble bed breeding
  blanket: Design status at the conclusion of the pre-concept design phase,
  Energies 16~(14) (2023) 5377.
\newblock \href {https://doi.org/10.3390/en16145377}
  {\path{doi:10.3390/en16145377}}.

\bibitem{Nakamichi-2018-NuclMatEne}
M.~Nakamichi, K.~J. Hwan, P.~Kurinskiy, M.~Nakamura, Thermal properties of
  beryllides as advanced neutron multipliers for {DEMO} fusion application,
  Nuclear Materials and Energy 15 (2018) 71--75.
\newblock \href {https://doi.org/10.1016/j.nme.2018.02.002}
  {\path{doi:10.1016/j.nme.2018.02.002}}.

\bibitem{Gohar-1980-ANS}
Y.~Gohar, M.~A. Abdou,
  \href{https://inis.iaea.org/search/search.aspx?orig_q=RN:12630669}{Neutronic
  optimization of solid breeder blankets for {STARFIRE} design}, Tech. Rep.
  CONF-801011-98, Argonne National Laboratory, Argonne, IL, USA (January 1980).
\newline\urlprefix\url{https://inis.iaea.org/search/search.aspx?orig_q=RN:12630669}

\bibitem{Donne-1986-JNuclMater}
M.~{Dalle Donne}, S.~Dorner, D.~F. Lupton, Fabrication and properties of
  {Zr\textsubscript{5}Pb\textsubscript{3}}, a new neutron multiplier material
  for fusion blankets, Journal of Nuclear Materials 141-143~(1) (1986)
  369--372.
\newblock \href {https://doi.org/10.1016/S0022-3115(86)80067-8}
  {\path{doi:10.1016/S0022-3115(86)80067-8}}.

\bibitem{Gaisin-2023-JMatResTech}
R.~Gaisin, P.~Pereslavtsev, S.~Baumgaertner, L.~Seemann, E.~Gaisina, V.~Chakin,
  S.~Udartsev, P.~Vladimirov, B.~Gorr, Lanthanum plumbide as a new neutron
  multiplier material, Journal of Materials Research and Technology 24 (2023)
  3399--3412.
\newblock \href {https://doi.org/10.1016/j.jmrt.2023.03.211}
  {\path{doi:10.1016/j.jmrt.2023.03.211}}.

\bibitem{Gao-2024-JNuclMater}
X.~Gao, J.~Wang, W.~Lu, Y.~Lu, D.~Chu, W.~Wang, Lithium lead titanate
  {(Li\textsubscript{2}Pb\textsubscript{x}Ti\textsubscript{1-x}O\textsubscript{3},
  0.1$<$x$<$0.9)}: A new tritium-neutron complex breeder for fusion reactor
  blanket, Journal of Nuclear Materials 599 (2024) 155240.
\newblock \href {https://doi.org/10.1016/j.jnucmat.2024.155240}
  {\path{doi:10.1016/j.jnucmat.2024.155240}}.

\bibitem{Rosenbluth-1997-NuclFus}
M.~N. Rosenbluth, S.~V. Putvinski, Theory for avalanche of runaway electrons in
  tokamaks, Nuclear Fusion 37~(10) (1997) 1355--1362.
\newblock \href {https://doi.org/10.1088/0029-5515/37/10/I03}
  {\path{doi:10.1088/0029-5515/37/10/I03}}.

\bibitem{Hesslow-2019-NuclFus}
L.~Hesslow, O.~Embr{\'e}us, O.~Vallhagen, T.~F{\"u}l{\"o}p, Influence of
  massive material injection on avalanche runaway generation during tokamak
  disruptions, Nuclear Fusion 59~(8) (2019) 084004.
\newblock \href {https://doi.org/10.1088/1741-4326/ab26c2}
  {\path{doi:10.1088/1741-4326/ab26c2}}.

\bibitem{Martin-Solis-2015-PhysPlas}
J.~Mart\'in-Sol\'is, A.~Loarte, M.~Lehnen, Runaway electron dynamics in tokamak
  plasmas with high impurity content, Physics of Plasmas 22~(9) (2015) 092512.
\newblock \href {https://doi.org/10.1063/1.4931166}
  {\path{doi:10.1063/1.4931166}}.

\bibitem{Hobbs-1967-PlasPhys}
G.~D. Hobbs, J.~A. Wesson, Heat flow through a {Langmuir} sheath in the
  presence of electron emission, Plasma Physics 9~(1) (1967) 85--87.
\newblock \href {https://doi.org/10.1088/0032-1028/9/1/410}
  {\path{doi:10.1088/0032-1028/9/1/410}}.

\bibitem{Chodura-1982-PhysFluid}
R.~Chodura, Plasma-wall transition in an oblique magnetic field, Physics of
  Fluids 25 (1982) 1628--1633.
\newblock \href {https://doi.org/10.1063/1.863955}
  {\path{doi:10.1063/1.863955}}.

\bibitem{Stangeby-2012-NuclFus}
P.~C. Stangeby, The {Chodura} sheath for angles of a few degrees between the
  magnetic field and the surface of divertor targets and limiters, Nuclear
  Fusion 52~(8) (2012) 083012.
\newblock \href {https://doi.org/10.1088/0029-5515/52/8/083012}
  {\path{doi:10.1088/0029-5515/52/8/083012}}.

\bibitem{Garland-2020-PhysPlas}
N.~A. Garland, H.-K. Chung, C.~J. Fontes, M.~C. Zammit, J.~Colgan, T.~Elder,
  C.~J. McDevitt, T.~M. Wildey, X.-Z. Tang, Impact of a minority relativistic
  electron tail interacting with a thermal plasma containing high-atomic-number
  impurities, Physics of Plasmas 27~(4) (2020) 040702.
\newblock \href {https://doi.org/10.1063/5.0003638}
  {\path{doi:10.1063/5.0003638}}.

\end{thebibliography}

%
\clearpage

\renewcommand{\theequation}{S\arabic{equation}}
\renewcommand{\thefigure}{S\arabic{figure}}

\renewcommand\figurename{Supplementary Figure}

\setcounter{ead}{0}
\setcounter{tnote}{0}
\setcounter{fnote}{0}
\setcounter{cnote}{0}
\setcounter{author}{0}
\setcounter{affn}{0}
\resetTitleCounters

\setcounter{figure}{0}

\makeatletter
\let\@title\@empty
\makeatother

\title{Supplementary data for the article entitled:\\
       Large radiation back-flux from Monte Carlo simulations
       of fusion neutron-material interactions}

\begin{abstract}
    In this supplement, we provide plots of all major data sets obtained
    during this work, which are tabulated in an abbreviated format in the
    main manuscript.
    The plotted data include energy-resolved distributions of the total
    radiation back-fluxes, operating-time-resolved integrated delayed
    back-fluxes, and some relevant neutron cross sections which are useful
    to interpret these data.
    These plots will be useful and informative to readers who seek greater
    detail on some points or those who wish to replicate or extend this
    work.
\end{abstract}

\maketitle

%

\clearpage

\section*{Total radiation back-fluxes}

\begin{figure}[h]
    \centering
    \begin{subfigure}{0.475\linewidth}
        \includegraphics[width=\linewidth]{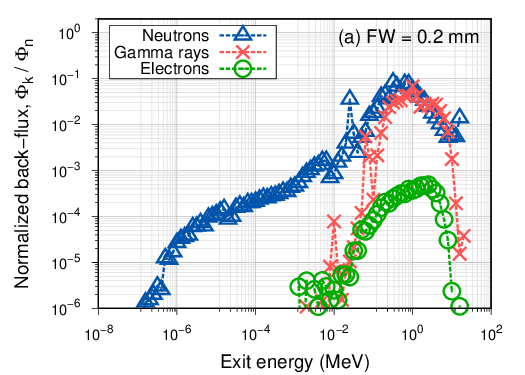}
        \centering
        \phantomcaption
        \label{subfig:erg-iron0.2}
    \end{subfigure}
    \begin{subfigure}{0.475\linewidth}
        \includegraphics[width=\linewidth]{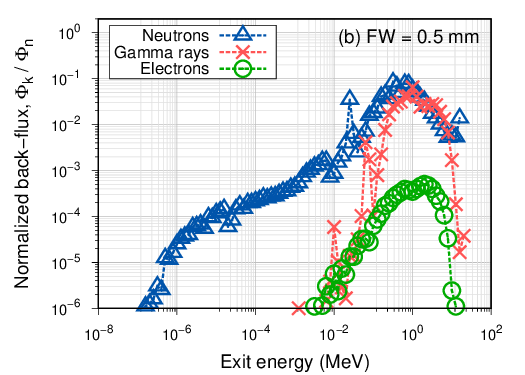}
        \centering
        \phantomcaption
        \label{subfig:erg-iron0.5}
    \end{subfigure}
    \begin{subfigure}{0.475\linewidth}
        \includegraphics[width=\linewidth]{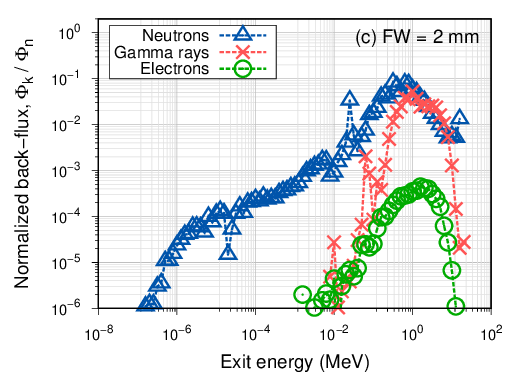}
        \centering
        \phantomcaption
        \label{subfig:erg-iron2}
    \end{subfigure}
    \begin{subfigure}{0.475\linewidth}
        \includegraphics[width=\linewidth]{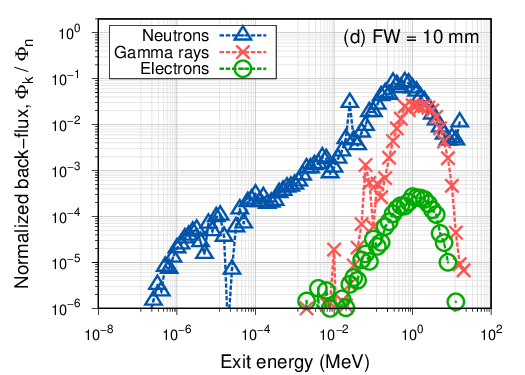}
        \centering
        \phantomcaption
        \label{subfig:erg-iron10}
    \end{subfigure}
    \caption{Energy-resolved, time-integrated radiation back-fluxes of
        neutrons, gamma rays, and relativistic electrons for iron
        structural material with tungsten first wall thicknesses of
        (\subref{subfig:erg-iron0.2}) 0.2 mm, (\subref{subfig:erg-iron0.5})
        0.5 mm, (\subref{subfig:erg-iron2}) 2 mm, and
        (\subref{subfig:erg-iron10}) 10 mm.}
    \label{fig:erg-iron}
\end{figure}

\begin{figure}[h]
    \centering
    \begin{subfigure}{0.475\linewidth}
        \includegraphics[width=\linewidth]{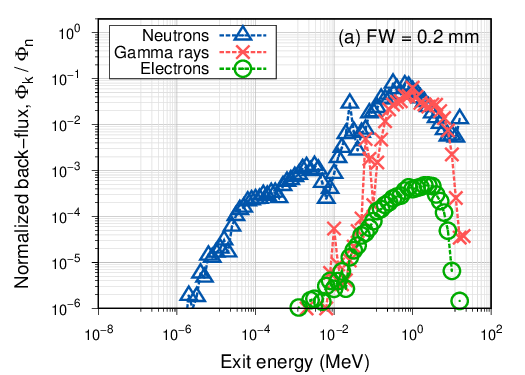}
        \centering
        \phantomcaption
        \label{subfig:s-erg-rafm0.2}
    \end{subfigure}
    \begin{subfigure}{0.475\linewidth}
        \includegraphics[width=\linewidth]{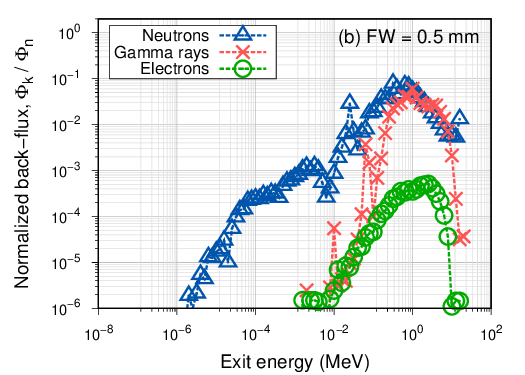}
        \centering
        \phantomcaption
        \label{subfig:s-erg-rafm0.5}
    \end{subfigure}
    \begin{subfigure}{0.475\linewidth}
        \includegraphics[width=\linewidth]{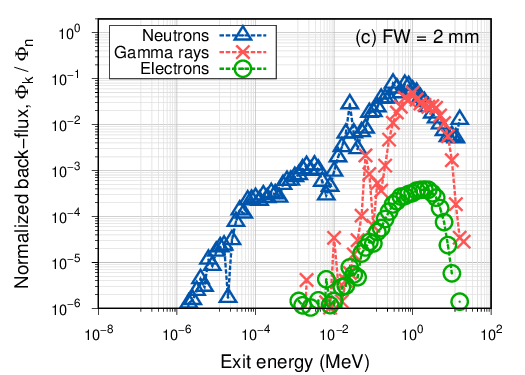}
        \centering
        \phantomcaption
        \label{subfig:s-erg-rafm2}
    \end{subfigure}
    \begin{subfigure}{0.475\linewidth}
        \includegraphics[width=\linewidth]{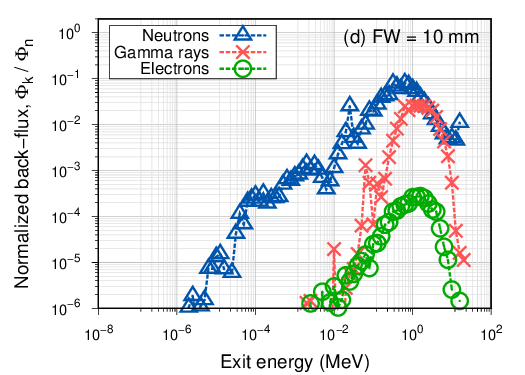}
        \centering
        \phantomcaption
        \label{subfig:s-erg-rafm10}
    \end{subfigure}
    \caption{Energy-resolved, time-integrated radiation back-fluxes of
        neutrons, gamma rays, and relativistic electrons for RAFM steel
        structural material with tungsten first wall thicknesses of
        (\subref{subfig:s-erg-rafm0.2}) 0.2 mm, (\subref{subfig:s-erg-rafm0.5})
        0.5 mm, (\subref{subfig:s-erg-rafm2}) 2 mm, and
        (\subref{subfig:s-erg-rafm10}) 10 mm.}
    \label{fig:s-erg-rafm}
\end{figure}

\begin{figure}[h]
    \centering
    \begin{subfigure}{0.475\linewidth}
        \includegraphics[width=\linewidth]{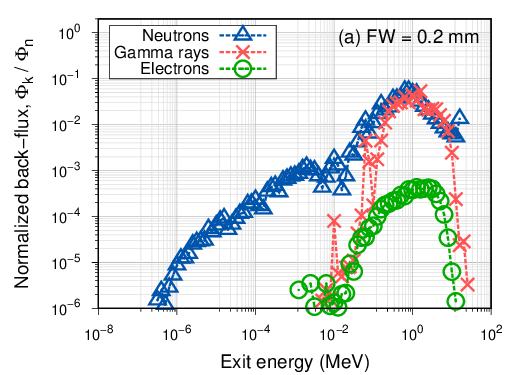}
        \centering
        \phantomcaption
        \label{subfig:erg-inc7180.2}
    \end{subfigure}
    \begin{subfigure}{0.475\linewidth}
        \includegraphics[width=\linewidth]{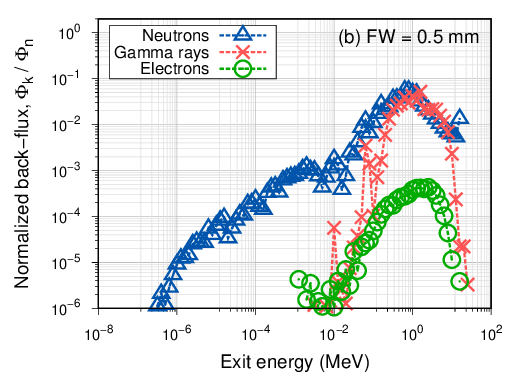}
        \centering
        \phantomcaption
        \label{subfig:erg-inc7180.5}
    \end{subfigure}
    \begin{subfigure}{0.475\linewidth}
        \includegraphics[width=\linewidth]{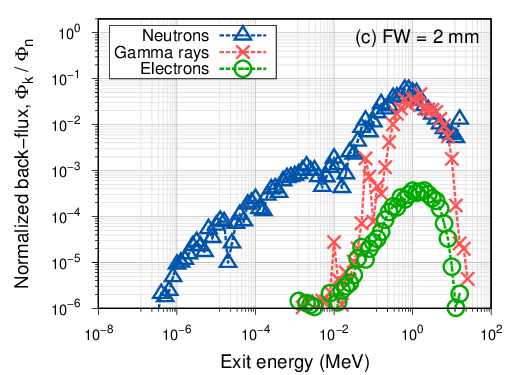}
        \centering
        \phantomcaption
        \label{subfig:erg-inc7182}
    \end{subfigure}
    \begin{subfigure}{0.475\linewidth}
        \includegraphics[width=\linewidth]{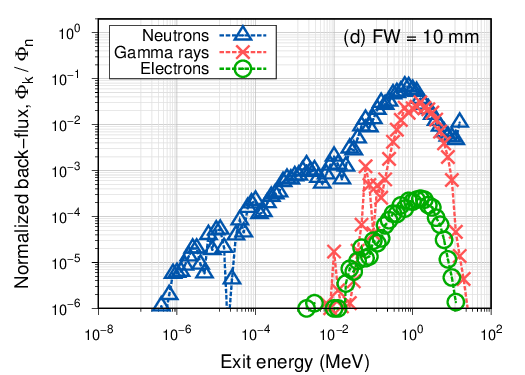}
        \centering
        \phantomcaption
        \label{subfig:erg-inc71810}
    \end{subfigure}
    \caption{Energy-resolved, time-integrated radiation back-fluxes of
        neutrons, gamma rays, and relativistic electrons for Inconel 718
        structural material with tungsten first wall thicknesses of
        (\subref{subfig:erg-inc7180.2}) 0.2 mm,
        (\subref{subfig:erg-inc7180.5}) 0.5 mm, (\subref{subfig:erg-inc7182})
        2 mm, and (\subref{subfig:erg-inc71810}) 10 mm.}
    \label{fig:erg-inc718}
\end{figure}

\begin{figure}[h]
    \centering
    \begin{subfigure}{0.475\linewidth}
        \includegraphics[width=\linewidth]{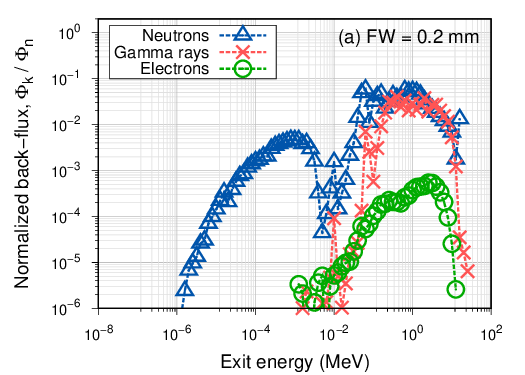}
        \centering
        \phantomcaption
        \label{subfig:erg-valloy0.2}
    \end{subfigure}
    \begin{subfigure}{0.475\linewidth}
        \includegraphics[width=\linewidth]{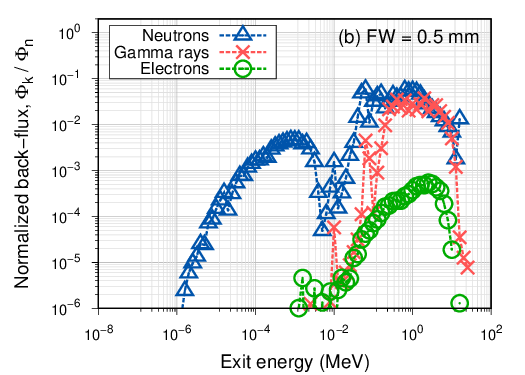}
        \centering
        \phantomcaption
        \label{subfig:erg-valloy0.5}
    \end{subfigure}
    \begin{subfigure}{0.475\linewidth}
        \includegraphics[width=\linewidth]{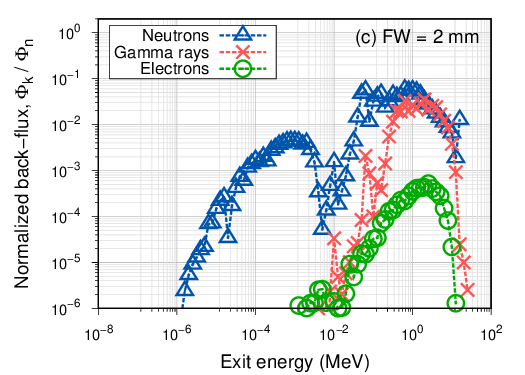}
        \centering
        \phantomcaption
        \label{subfig:erg-valloy2}
    \end{subfigure}
    \begin{subfigure}{0.475\linewidth}
        \includegraphics[width=\linewidth]{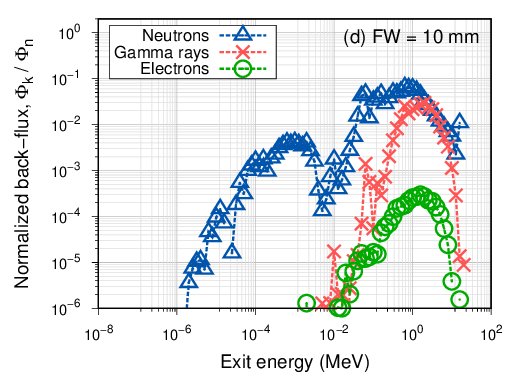}
        \centering
        \phantomcaption
        \label{subfig:erg-valloy10}
    \end{subfigure}
    \caption{Energy-resolved, time-integrated radiation back-fluxes of
        neutrons, gamma rays, and relativistic electrons for V-4Ti-4Cr
        structural material with tungsten first wall thicknesses of
        (\subref{subfig:erg-valloy0.2}) 0.2 mm,
        (\subref{subfig:erg-valloy0.5}) 0.5 mm, (\subref{subfig:erg-valloy2})
        2 mm, and (\subref{subfig:erg-valloy10}) 10 mm.}
    \label{fig:erg-valloy}
\end{figure}

Energy distributions for neutron, gamma ray, and electron back-fluxes are
plotted in
\cref{fig:erg-iron,fig:s-erg-rafm,fig:erg-inc718,fig:erg-valloy} for
all material configurations studied in this work.

\clearpage

\begin{figure}[h]
    \centering
    \begin{subfigure}{0.475\linewidth}
        \includegraphics[width=\linewidth]{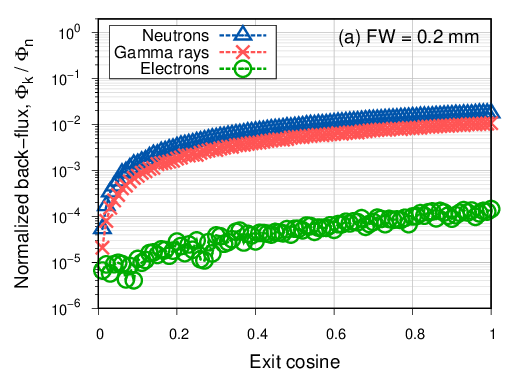}
        \centering
        \phantomcaption
        \label{subfig:cos-iron0.2}
    \end{subfigure}
    \begin{subfigure}{0.475\linewidth}
        \includegraphics[width=\linewidth]{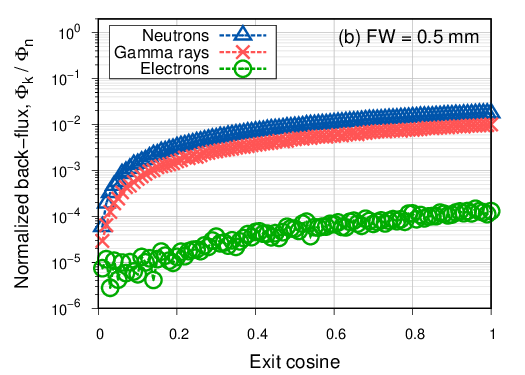}
        \centering
        \phantomcaption
        \label{subfig:cos-iron0.5}
    \end{subfigure}
    \begin{subfigure}{0.475\linewidth}
        \includegraphics[width=\linewidth]{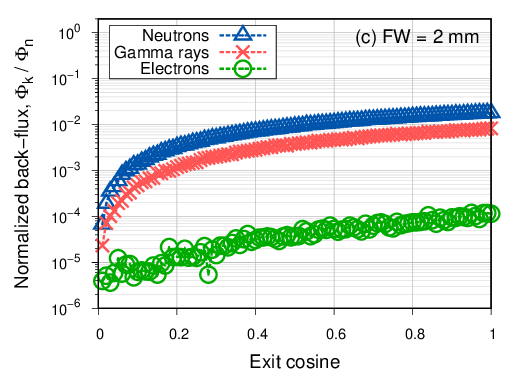}
        \centering
        \phantomcaption
        \label{subfig:cos-iron2}
    \end{subfigure}
    \begin{subfigure}{0.475\linewidth}
        \includegraphics[width=\linewidth]{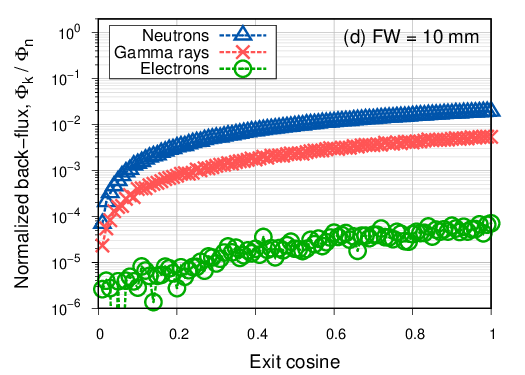}
        \centering
        \phantomcaption
        \label{subfig:cos-iron10}
    \end{subfigure}
    \caption{Cosine-resolved, time-integrated radiation back-fluxes of
        neutrons, gamma rays, and relativistic electrons for iron
        structural material with tungsten first wall thicknesses of
        (\subref{subfig:cos-iron0.2}) 0.2 mm, (\subref{subfig:cos-iron0.5})
        0.5 mm, (\subref{subfig:cos-iron2}) 2 mm, and
        (\subref{subfig:cos-iron10}) 10 mm.}
    \label{fig:cos-iron}
\end{figure}

\begin{figure}[h]
    \centering
    \begin{subfigure}{0.475\linewidth}
        \includegraphics[width=\linewidth]{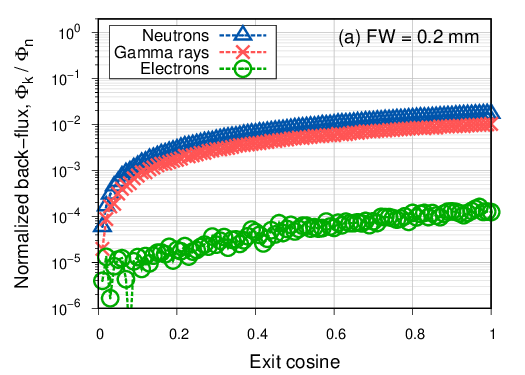}
        \centering
        \phantomcaption
        \label{subfig:cos-rafm0.2}
    \end{subfigure}
    \begin{subfigure}{0.475\linewidth}
        \includegraphics[width=\linewidth]{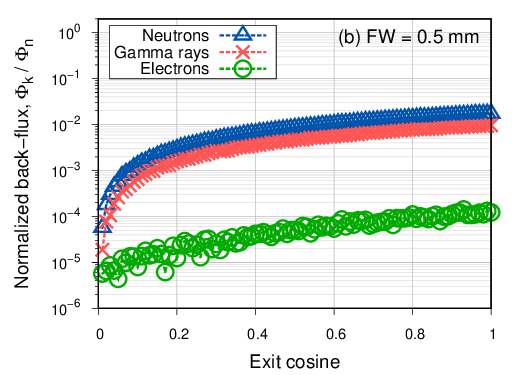}
        \centering
        \phantomcaption
        \label{subfig:cos-rafm0.5}
    \end{subfigure}
    \begin{subfigure}{0.475\linewidth}
        \includegraphics[width=\linewidth]{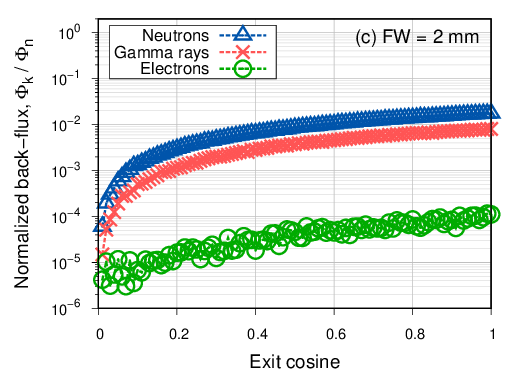}
        \centering
        \phantomcaption
        \label{subfig:cos-rafm2}
    \end{subfigure}
    \begin{subfigure}{0.475\linewidth}
        \includegraphics[width=\linewidth]{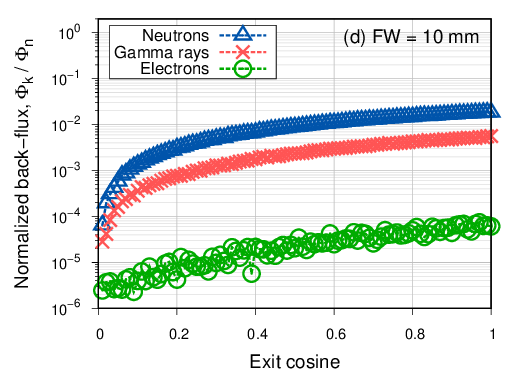}
        \centering
        \phantomcaption
        \label{subfig:cos-rafm10}
    \end{subfigure}
    \caption{Cosine-resolved, time-integrated radiation back-fluxes of
        neutrons, gamma rays, and relativistic electrons for RAFM steel
        structural material with tungsten first wall thicknesses of
        (\subref{subfig:cos-rafm0.2}) 0.2 mm, (\subref{subfig:cos-rafm0.5})
        0.5 mm, (\subref{subfig:cos-rafm2}) 2 mm, and
        (\subref{subfig:cos-rafm10}) 10 mm.}
    \label{fig:cos-rafm}
\end{figure}

\begin{figure}[h]
    \centering
    \begin{subfigure}{0.475\linewidth}
        \includegraphics[width=\linewidth]{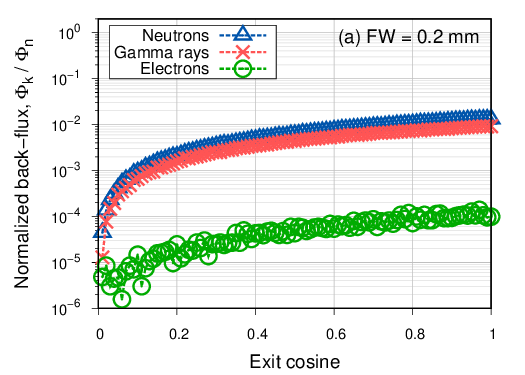}
        \centering
        \phantomcaption
        \label{subfig:cos-inc7180.2}
    \end{subfigure}
    \begin{subfigure}{0.475\linewidth}
        \includegraphics[width=\linewidth]{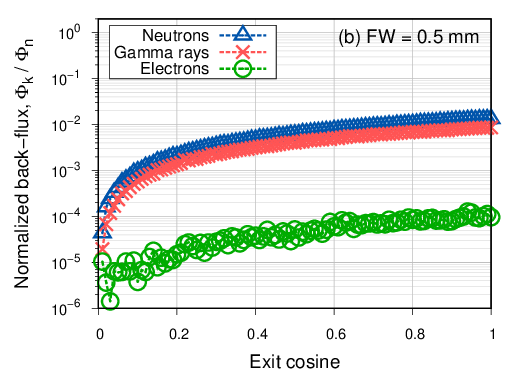}
        \centering
        \phantomcaption
        \label{subfig:cos-inc7180.5}
    \end{subfigure}
    \begin{subfigure}{0.475\linewidth}
        \includegraphics[width=\linewidth]{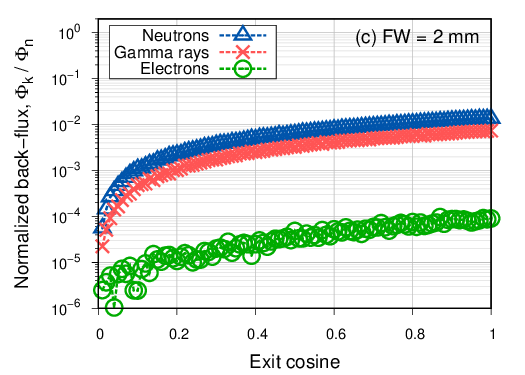}
        \centering
        \phantomcaption
        \label{subfig:cos-inc7182}
    \end{subfigure}
    \begin{subfigure}{0.475\linewidth}
        \includegraphics[width=\linewidth]{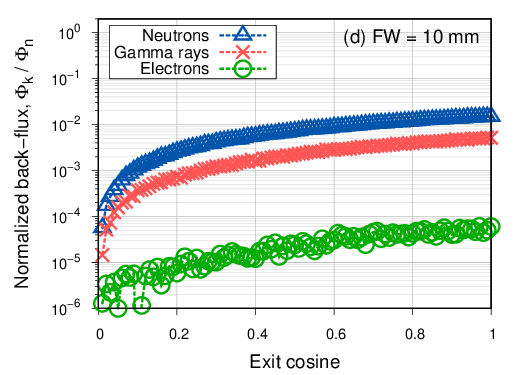}
        \centering
        \phantomcaption
        \label{subfig:cos-inc71810}
    \end{subfigure}
    \caption{Cosine-resolved, time-integrated radiation back-fluxes of
        neutrons, gamma rays, and relativistic electrons for Inconel 718
        structural material with tungsten first wall thicknesses of
        (\subref{subfig:cos-inc7180.2}) 0.2 mm,
        (\subref{subfig:cos-inc7180.5}) 0.5 mm, (\subref{subfig:cos-inc7182})
        2 mm, and (\subref{subfig:cos-inc71810}) 10 mm.}
    \label{fig:cos-inc718}
\end{figure}

\begin{figure}[h]
    \centering
    \begin{subfigure}{0.475\linewidth}
        \includegraphics[width=\linewidth]{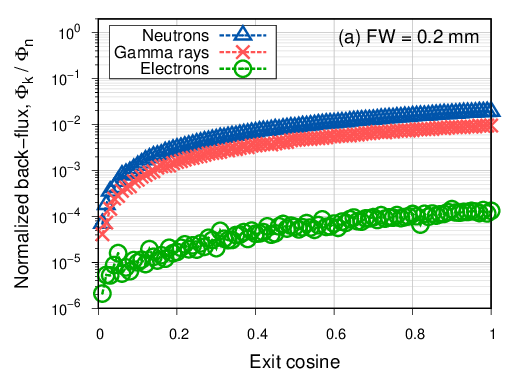}
        \centering
        \phantomcaption
        \label{subfig:cos-valloy0.2}
    \end{subfigure}
    \begin{subfigure}{0.475\linewidth}
        \includegraphics[width=\linewidth]{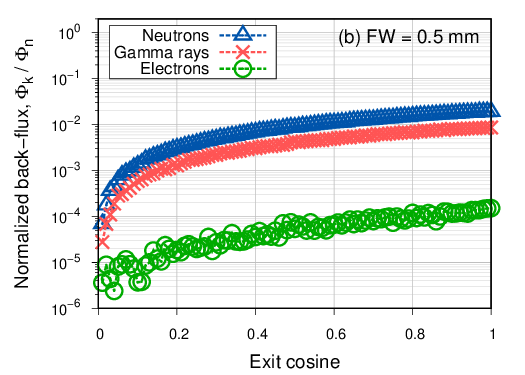}
        \centering
        \phantomcaption
        \label{subfig:cos-valloy0.5}
    \end{subfigure}
    \begin{subfigure}{0.475\linewidth}
        \includegraphics[width=\linewidth]{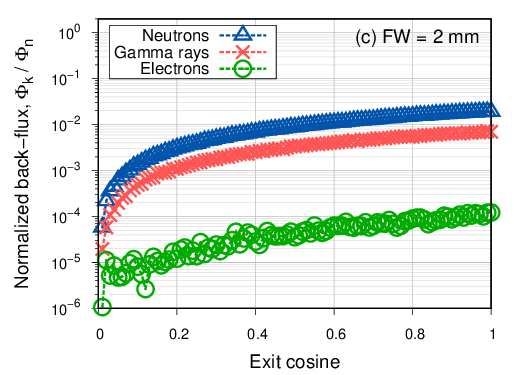}
        \centering
        \phantomcaption
        \label{subfig:cos-valloy2}
    \end{subfigure}
    \begin{subfigure}{0.475\linewidth}
        \includegraphics[width=\linewidth]{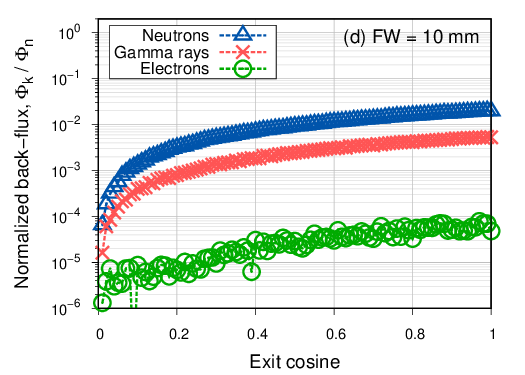}
        \centering
        \phantomcaption
        \label{subfig:cos-valloy10}
    \end{subfigure}
    \caption{Cosine-resolved, time-integrated radiation back-fluxes of
        neutrons, gamma rays, and relativistic electrons for V-4Ti-4Cr
        structural material with tungsten first wall thicknesses of
        (\subref{subfig:cos-valloy0.2}) 0.2 mm,
        (\subref{subfig:cos-valloy0.5}) 0.5 mm, (\subref{subfig:cos-valloy2})
        2 mm, and (\subref{subfig:cos-valloy10}) 10 mm.}
    \label{fig:cos-valloy}
\end{figure}

Cosine distributions for neutron, gamma ray, and electron back-fluxes are
plotted in \cref{fig:cos-iron,fig:cos-rafm,fig:cos-inc718,fig:cos-valloy} for
all material configurations studied in this work.

\clearpage

\section*{Time-integrated delayed back-fluxes}

\begin{figure}[h]
    \centering
    \begin{subfigure}{0.475\linewidth}
        \centering
        \phantomcaption
        \includegraphics[width=\linewidth]{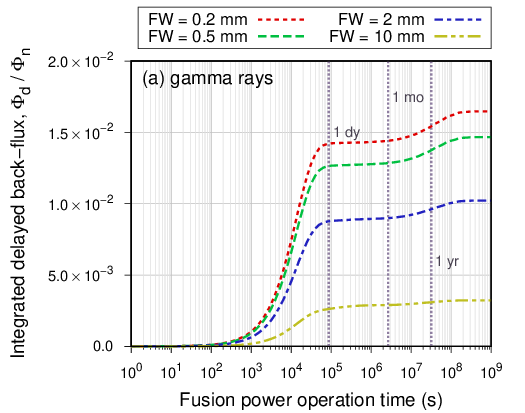}
        \label{subfig:delay-ph-iron}
    \end{subfigure}
    \begin{subfigure}{0.475\linewidth}
        \centering
        \phantomcaption
        \includegraphics[width=\linewidth]{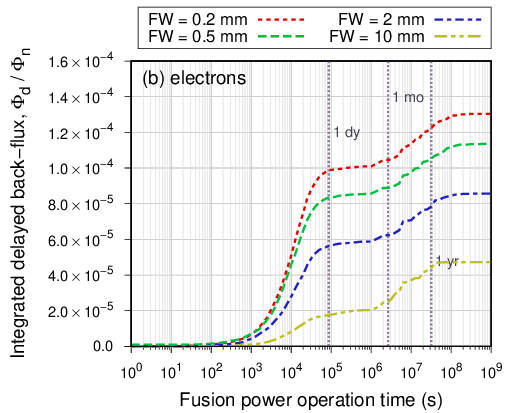}
        \label{subfig:delay-el-iron}
    \end{subfigure}
    \caption{Time-integrated delayed back-fluxes of
        (\subref{subfig:delay-ph-iron}) gamma rays and
        (\subref{subfig:delay-el-iron}) electrons for different first wall
        thicknesses with iron structural material.}
    \label{fig:delay-iron}
\end{figure}

\begin{figure}[h]
    \centering
    \begin{subfigure}{0.475\linewidth}
        \centering
        \phantomcaption
        \includegraphics[width=\linewidth]{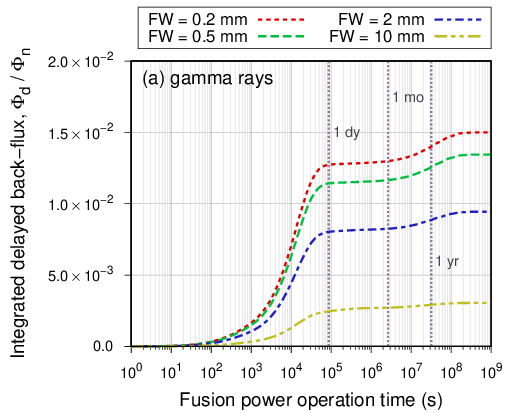}
        \label{subfig:delay-ph-rafm}
    \end{subfigure}
    \begin{subfigure}{0.475\linewidth}
        \centering
        \phantomcaption
        \includegraphics[width=\linewidth]{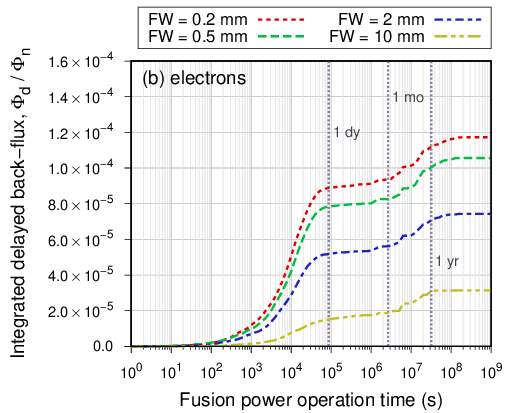}
        \label{subfig:delay-el-rafm}
    \end{subfigure}
    \caption{Time-integrated delayed back-fluxes of
        (\subref{subfig:delay-ph-rafm}) gamma rays and
        (\subref{subfig:delay-el-rafm}) electrons for different first wall
        thicknesses with RAFM steel structural material.}
    \label{fig:delay-rafm}
\end{figure}

\begin{figure}[h]
    \centering
    \begin{subfigure}{0.475\linewidth}
        \centering
        \phantomcaption
        \includegraphics[width=\linewidth]{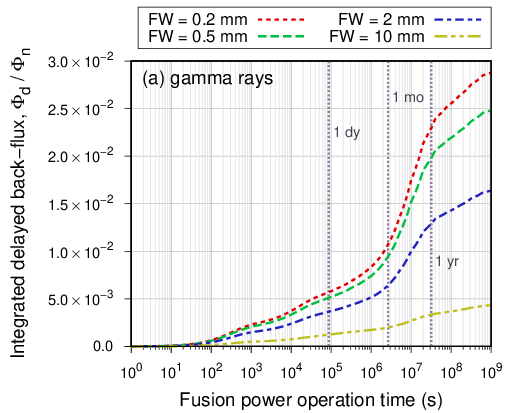}
        \label{subfig:delay-ph-inc718}
    \end{subfigure}
    \begin{subfigure}{0.475\linewidth}
        \centering
        \phantomcaption
        \includegraphics[width=\linewidth]{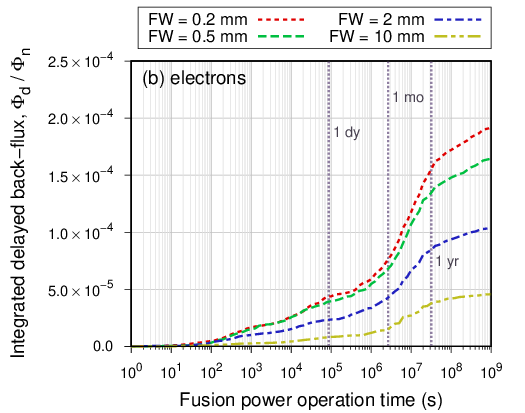}
        \label{subfig:delay-el-inc718}
    \end{subfigure}
    \caption{Time-integrated delayed back-fluxes of
        (\subref{subfig:delay-ph-inc718}) gamma rays and
        (\subref{subfig:delay-el-inc718}) electrons for different first wall
        thicknesses with Inconel 718 structural material.}
    \label{fig:delay-inc718}
\end{figure}

\begin{figure}[h]
    \centering
    \begin{subfigure}{0.475\linewidth}
        \centering
        \phantomcaption
        \includegraphics[width=\linewidth]{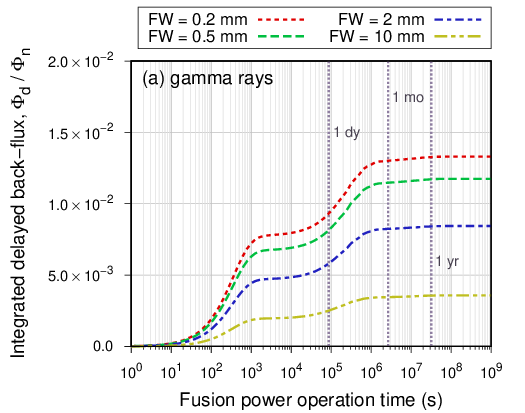}
        \label{subfig:delay-ph-valloy}
    \end{subfigure}
    \begin{subfigure}{0.475\linewidth}
        \centering
        \phantomcaption
        \includegraphics[width=\linewidth]{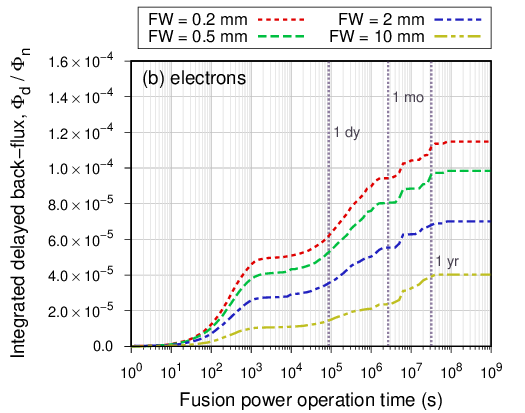}
        \label{subfig:delay-el-valloy}
    \end{subfigure}
    \caption{Time-integrated delayed back-fluxes of
        (\subref{subfig:delay-ph-valloy}) gamma rays and
        (\subref{subfig:delay-el-valloy}) electrons for different first wall
        thicknesses with V-4Ti-4Cr structural material.}
    \label{fig:delay-valloy}
\end{figure}

Time-integrated delayed back-fluxes for gamma rays and electrons are
plotted in
\cref{fig:delay-iron,fig:delay-rafm,fig:delay-inc718,fig:delay-valloy} for
all material configurations studied in this work.

\clearpage

\section*{Neutron cross sections}

The following relevant cross sections are plotted here:
\begin{itemize}
    \item Total cross section, $\sigma_\mathrm{t}$: the total probability for
    a neutron to interact with an atom of the material.
    \item Neutron multiplication cross section, $\sigma_\mathrm{m}$: the
    weighted probability for a neutron collision to generate additional
    neutrons, evaluated as
    \begin{equation}
        \sigma_\mathrm{m} = \sum_{k=1}^{N_k} X_k \sigma_k
    \end{equation}
    where $\sigma_k$ is a reaction which might produce secondary neutrons and
    $X_k$ is the number of secondary neutrons produced, which may be
    energy-dependent. In general, a larger ratio $\sigma_\mathrm{m} /
    \sigma_\mathrm{t}$ indicates that a neutron in a material is more likely
    to generate secondary neutrons before losing energy or being captured.
    \item Secondary neutron energy probability distribution functions (PDFs)
    at various incident neutron energies above the threshold for
    $\sigma_\mathrm{m}$.
\end{itemize}
Cross sections and secondary neutron energy PDFs for a given material are the
weighted sum of the per-isotope cross sections and PDFs. These data were
extracted from the ENDF/B-VIII.0 library in ACE format used by MCNP with the
ACEtk package (\url{https://github.com/njoy/ACEtk}).

\begin{figure}[h]
    \centering
    \begin{subfigure}{0.475\linewidth}
        \includegraphics[width=\linewidth]{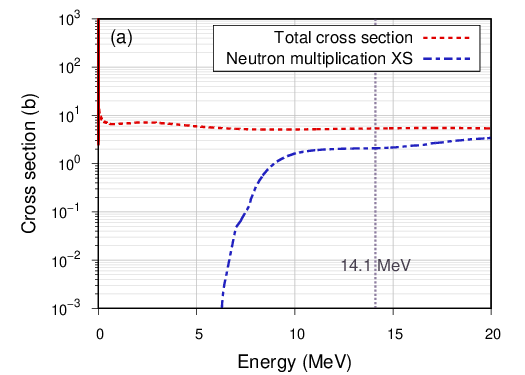}
        \centering
        \phantomcaption
        \label{subfig:tungsten-xs}
    \end{subfigure}
    \begin{subfigure}{0.475\linewidth}
        \includegraphics[width=\linewidth]{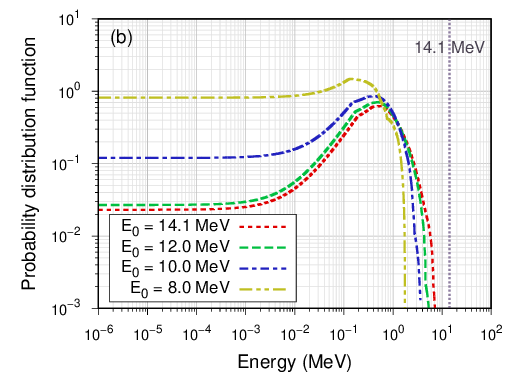}
        \centering
        \phantomcaption
        \label{subfig:tungsten-sec-erg}
    \end{subfigure}
    \caption{Neutron cross section data for tungsten:
        (\subref{subfig:tungsten-xs})
        Total and neutron multiplication cross sections.
        (\subref{subfig:tungsten-sec-erg}) Secondary neutron energy
        probability
        distribution functions for various incident neutron energies. Dashed
        vertical lines indicate the 14.1 MeV fusion neutron energy as a useful
        reference point.}
    \label{fig:tungsten-xsdata}
\end{figure}

\begin{figure}[h]
    \centering
    \begin{subfigure}{0.475\linewidth}
        \includegraphics[width=\linewidth]{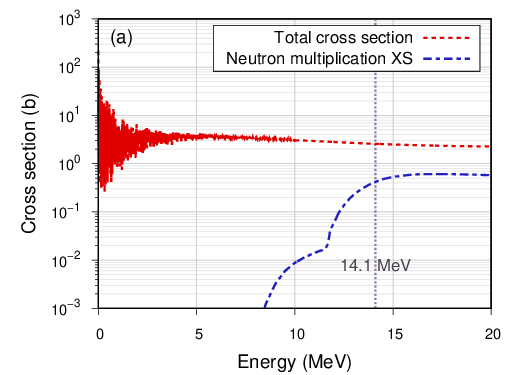}
        \centering
        \phantomcaption
        \label{subfig:iron-xs}
    \end{subfigure}
    \begin{subfigure}{0.475\linewidth}
        \includegraphics[width=\linewidth]{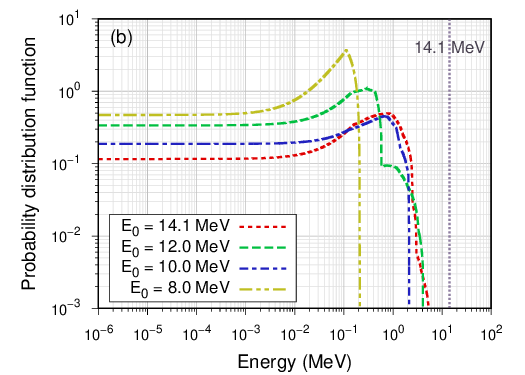}
        \centering
        \phantomcaption
        \label{subfig:iron-sec-erg}
    \end{subfigure}
    \caption{Neutron cross section data for iron: (\subref{subfig:iron-xs})
        Total and neutron multiplication cross sections.
        (\subref{subfig:iron-sec-erg}) Secondary neutron energy probability
        distribution functions for various incident neutron energies. Dashed
        vertical lines indicate the 14.1 MeV fusion neutron energy as a useful
        reference point.}
    \label{fig:iron-xsdata}
\end{figure}

\begin{figure}[h]
    \centering
    \begin{subfigure}{0.475\linewidth}
        \includegraphics[width=\linewidth]{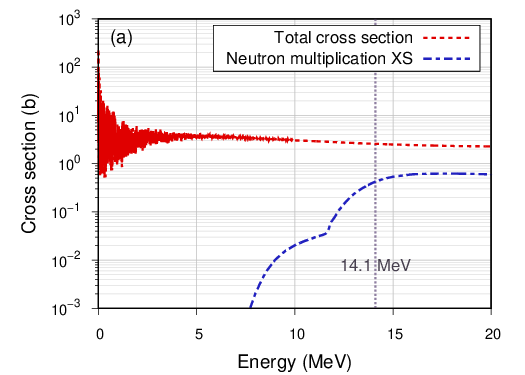}
        \centering
        \phantomcaption
        \label{subfig:rafm-xs}
    \end{subfigure}
    \begin{subfigure}{0.475\linewidth}
        \includegraphics[width=\linewidth]{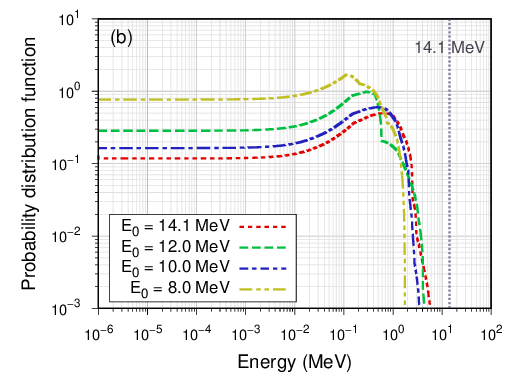}
        \centering
        \phantomcaption
        \label{subfig:rafm-sec-erg}
    \end{subfigure}
    \caption{Neutron cross section data for RAFM steel:
        (\subref{subfig:rafm-xs})
        Total and neutron multiplication cross sections.
        (\subref{subfig:rafm-sec-erg}) Secondary neutron energy probability
        distribution functions for various incident neutron energies. Dashed
        vertical lines indicate the 14.1 MeV fusion neutron energy as a useful
        reference point.}
    \label{fig:rafm-xsdata}
\end{figure}

\begin{figure}[h]
    \centering
    \begin{subfigure}{0.475\linewidth}
        \includegraphics[width=\linewidth]{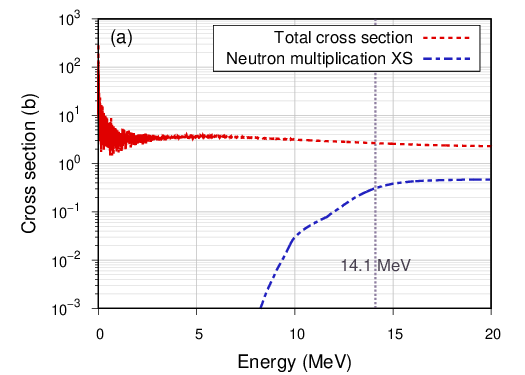}
        \centering
        \phantomcaption
        \label{subfig:inc718-xs}
    \end{subfigure}
    \begin{subfigure}{0.475\linewidth}
        \includegraphics[width=\linewidth]{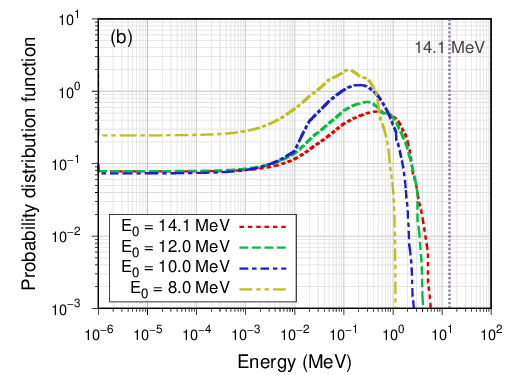}
        \centering
        \phantomcaption
        \label{subfig:inc718-sec-erg}
    \end{subfigure}
    \caption{Neutron cross section data for Inconel 718:
        (\subref{subfig:inc718-xs})
        Total and neutron multiplication cross sections.
        (\subref{subfig:inc718-sec-erg}) Secondary neutron energy probability
        distribution functions for various incident neutron energies. Dashed
        vertical lines indicate the 14.1 MeV fusion neutron energy as a useful
        reference point.}
    \label{fig:inc718-xsdata}
\end{figure}

\begin{figure}[h]
    \centering
    \begin{subfigure}{0.475\linewidth}
        \includegraphics[width=\linewidth]{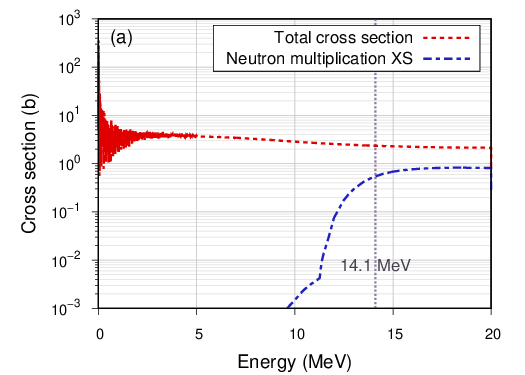}
        \centering
        \phantomcaption
        \label{subfig:valloy-xs}
    \end{subfigure}
    \begin{subfigure}{0.475\linewidth}
        \includegraphics[width=\linewidth]{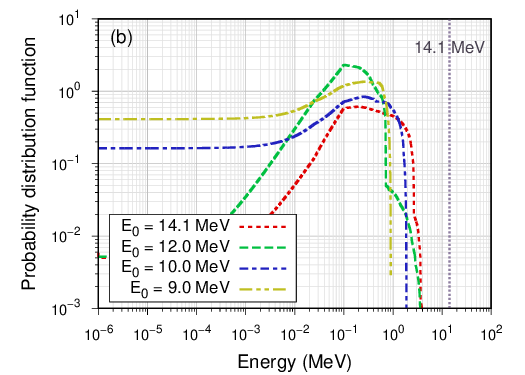}
        \centering
        \phantomcaption
        \label{subfig:valloy-sec-erg}
    \end{subfigure}
    \caption{Neutron cross section data for V-4Ti-4Cr:
        (\subref{subfig:valloy-xs})
        Total and neutron multiplication cross sections.
        (\subref{subfig:valloy-sec-erg}) Secondary neutron energy probability
        distribution functions for various incident neutron energies. Dashed
        vertical lines indicate the 14.1 MeV fusion neutron energy as a useful
        reference point.}
    \label{fig:valloy-xsdata}
\end{figure}

\end{document}